\documentclass[12pt]{iopart}
\usepackage{iopams}
\newcommand{\deq}{\stackrel{\mathrm{def}}{=}}
\def\lg{\langle}
\def\rg{\rangle}

\begin{document}
\title[Finite tight frames and some applications]{Finite tight frames and some applications}
\author{Nicolae Cotfas$^1$ and Jean Pierre Gazeau$^2$}
\address{$^1$Faculty of Physics, University of Bucharest, PO Box 76 - 54, Post Office 76, Bucharest, Romania}
\address{$^2$ Laboratoire APC, Universit\'e Paris 7-Denis Diderot, 10, rue A. Domon et L. Duquet, 75205 Paris Cedex13, France}
\eads{\mailto{ncotfas@yahoo.com}, \mailto{gazeau@apc.univ-paris7.fr}}
\begin{abstract} 
A finite-dimensional Hilbert space is usually described in terms of an orthonormal basis, but in certain approaches or applications
a description in terms of a finite overcomplete system of vectors, called a finite tight frame, may offer some advantages.
The use of a finite tight frame may lead to a simpler description of the symmetry transformations, to a simpler and
more symmetric form of invariants or to the possibility to define new mathematical objects with physical meaning, particularly
in regard with the notion of a quantization of a finite set.
We present some results concerning the use of integer coefficients and frame quantization, several examples and
suggest some possible applications. 
\end{abstract}
\maketitle
\section{Introduction}
Although, at first glance, a system described by a finite-dimensional Hilbert space looks much simpler
than one described by an infinite dimensional space, there is much more knowledge about the latter than the former.
The continuous systems of coherent states have many applications \cite{G2000,K1985,P1986} but the corresponding discrete
version, usually called a frame, seems to be less used in quantum physics. 
Hilbert space frames, introduced by Duffin and Schaeffer in their work on nonharmonic Fourier series \cite{Duffin}, 
were later rediscovered by Daubechies, Grossmann and Meyer in the fundamental paper \cite{DGM}.
Finite frames \cite{G2000,Balazs,Bene,Ch,G2009,Ge} are useful in  finite-dimensional  quantum mechanics \cite{Z}, particularly in quantum information \cite{Klya2007,Klya2008,badzi09},
and play a significant role in signal processing (they give stable signal representations and allow modeling 
for noisy environments) \cite{Daubechies}.
Our aim is to present some results concerning the finite frames and their applications in physics, particularly in the context of quantization of finite sets.
Particularly also, we try to prove that some mathematical methods used in modeling crystalline or quasicrystalline structures 
are in fact based on certain finite frames.

Each finite frame in a Hilbert space $\mathcal{H}$ defines an embedding of $\mathcal{H}$ into a higher dimensional
Hilbert space (called a superspace), and conversely, each embedding of $\mathcal{H}$ into a superspace
allows us to define some finite frames. The embedding into a superspace offers the possibility to define
some new mathematical objects, useful in certain applications. 
The construction of coherent states proposed by Perelomov in the case of Lie groups \cite{P1972} admits a version 
for finite groups, and leads to some useful finite frames. Certain representations in terms of finite frames can 
be regarded as Riemann sums corresponding to the integrals occurring in some representations in terms of 
continuous frames.

The description of a physical system in terms of a finite frame allows us to associate a linear operator
to a {\it classical} observable. The procedure, not necessarily a path to a quantum approach, can be regarded 
as an extended version of the Klauder-Berezin-Toeplitz quantization \cite{B,K1963a,K1963b,K1995} 
and represents a change of point of view in considering the physical 
system \cite{G2007a,G2009,G2004a,G2004b, G2006, G2007b,G2003}.

The paper is organized as follows. In section 2 we review some basic elements concerning the notion of tight frame in form suitable for the
applications in crystal physics and finite frame quantization we present throughout the paper. 
We explain how Parseval frames are easily constructed by projection from higher-dimensional spaces, and show  how a superspace emerges naturally from the existence of a frame in a given Hilbert space. By following the analogy with the systems of coherent states
we introduce the notion of normalized Parseval frame, define its proximity to an orthonormal basis in terms of a natural parameter $\eta$ and describe some stochastic aspects.
A Perelomov-like construction of frames through group representations is described at the end of the section.
By taking into consideration the embedding into superspace, we investigate in section 3 the set of the elements which can be represented 
as a linear combination with integer coefficients of the frame vectors, and present some applications. 
We show in which way some simple crystalline structures in the plane or in space are naturally described with the aid of frames. 
Section 4 is devoted to what we call frame quantization of discrete variable functions. Frame quantization replaces such functions by matrices, introducing in this way noncommutative algebras of matrices. We present an interesting result issued from  the stochastic aspects mentioned in section 2.  We also introduce another parameter, $\zeta$, expressing the distance of the ``quantum" non-commutative world issued from the frame quantization to the classical commutative one. 
We then illustrate our results concerning  the proximity of the ``quantum non-commutativity''   to the original ``classical'' commutativity when the number of elements of a frame is larger by one than the dimension of the vector space.
%
%
%
%
%
\section{Finite tight frames}
\subsection{Finite frames}
Let $\mathbb{K}$ be the field $\mathbb{R}$ or $\mathbb{C}$, and let  
$\mathcal{H}$ be a $N$-dimensional Hilbert space over $\mathbb{K}$ with
$\{ |j\rangle \}_{j=1}^N$ a fixed orthonormal basis.
A system of vectors $\{|w_i\rangle \}_{i=1}^M$ is a {\it finite frame} for $\mathcal{H}$  if 
there are constants $0<A \leq B <\infty $ such that 
\begin{equation} 
A ||v||^2\leq \sum_{i=1}^M|\langle w_i|v\rangle |^2\leq B ||v||^2\qquad 
{\rm for\ all}\quad |v\rangle \in \mathcal{H}.
\end{equation} 
The frame operator 
\begin{equation} 
S|v\rangle =\sum_{i=1}^M|w_i\rangle \langle w_i|v\rangle 
\end{equation} 
satisfies the relation
\[
\langle v|A v\rangle =A ||v||^2\leq \sum_{i=1}^M|\langle w_i|v\rangle |^2
= \langle v|Sv\rangle \leq B ||v||^2=\langle v|B v\rangle
\]
that is, 
\[
A \, \mathbb{I}_\mathcal{H}\leq S\leq B \, \mathbb{I}_\mathcal{H}
\]
where $\mathbb{I}_\mathcal{H}$ is the identity operator.
If $A =B $, the frame is called an $A $-{\it tight frame} and  
\[
S=A \, \mathbb{I}_\mathcal{H}.
\]
A frame $\{|w_i\rangle \}_{i=1}^M$ is called an {\it equal norm frame} if
$||w_1||=||w_2||=\cdots =||w_M||.$
A $1$-tight frame is usually called a {\em Parseval frame} and in this case  
\begin{equation}\label{rezu}
\sum_{i=1}^M|w_i\rangle \langle w_i |=\mathbb{I}_\mathcal{H}. 
\end{equation}
If $\{|w_i\rangle \}_{i=1}^M$ is an $A$-tight frame then  
$\{\frac{1}{\sqrt{A}}|w_i\rangle \}_{i=1}^M$ is a Parseval frame.
\subsection{Finite normalized Parseval frames}
Finite frames play a fundamental role in a wide variety of areas, and generally, each application requires a specific class of frames.
In the case of finite frame quantization, we regard a Parseval frame as a finite 
family of coherent states. In order to improve the correspondence between the two notions we consider Parseval frames which do not contain 
the null vector and express their vectors in terms of some unit vectors.

Let $\{|w_i\rangle \}_{i=1}^M$ be a Parseval frame. Denoting
\[
\kappa _i=\langle w_i|w_i\rangle \qquad {\rm and}\qquad |u_i\rangle =\frac{1}{\sqrt{\kappa _i}}\, |w_i\rangle 
\]
the resolution of identity (\ref{rezu}) becomes
\begin{equation}
\sum_{i=1}^M\kappa _i\, |u_i\rangle \langle u_i |=\mathbb{I}_\mathcal{H}. 
\end{equation}
We have
\begin{equation} \label{vscalarw}
\langle v|w\rangle =\sum_{i=1}^M \kappa _i\, \langle v|u_i\rangle \, \langle u_i|w\rangle ,\qquad 
||v||^2=\sum_{i=1}^M \kappa _i\, |\langle u_i|v\rangle |^2 
\end{equation}  
for any $|v\rangle ,\, |w\rangle \in \mathcal{H}$, and the well-known \cite{Go98,Go01,Zi} relation
\begin{equation} 
N=\sum_{j=1}^N\langle j|j\rangle =\sum_{j=1}^N\sum_{i=1}^M\kappa _i|\langle u_i|j\rangle |^2= 
\sum_{i=1}^M\kappa _i \sum_{j=1}^N|\langle u_i|j\rangle |^2=\sum_{i=1}^M\kappa _i .
\end{equation}
In this paper, by {\it normalized Parseval frame} in $\mathcal{H}$ we mean any system of vectors $\{ |u_i\rangle \}_{i=1}^M$ 
satisfying the following two conditions:\\
\qquad 1) the vectors $|u_i\rangle $ are unit vectors, that is,
\[
\langle u_i|u_i\rangle =1, \qquad {\rm for\ any\  } i\in \{ 1,2,\dots M\}
\]
\qquad 2) there are $\{\kappa _i\}_{i=1}^M$ positive constants such that
\begin{equation}\label{ff}
\sum_{i=1}^M\kappa _i\, |u_i\rangle \langle u_i |=\mathbb{I}_\mathcal{H}. 
\end{equation} 
If $\{ |u_i\rangle \}_{i=1}^M$ is a normalized Parseval frame with the constants $\{\kappa _i\}_{i=1}^M$ then $\{ \sqrt{\kappa _i}|u_i\rangle \}_{i=1}^M$ is
a Parseval frame, and conversely, if $\{ |w_i\rangle \}_{i=1}^M$ is a Parseval frame then $\{ \frac{1}{||w_i||}|w_i\rangle \}_{i=1}^M$
is a normalized Parseval frame with the constants $\{ ||w_i||^2 \}_{i=1}^M$.
In the case $\kappa _1=\kappa _2= ...=\kappa _M$, the relations (\ref{ff}) and (\ref{vscalarw}) become \cite{Go01,Go98,Zi}
\begin{equation}
\frac{N}{M}\sum_{i=1}^M |u_i\rangle \langle u_i|=\mathbb{I}_\mathcal{H}
\end{equation}
respectively
\begin{equation}\label{normm}
\langle v|w\rangle =\frac{N}{M}\sum_{i=1}^M \langle v|u_i\rangle \, \langle u_i|w\rangle ,\qquad 
||v||^2=\frac{N}{M}\sum_{i=1}^M |\langle u_i|v\rangle |^2 .
\end{equation}
and the  frame is called a {\em finite equal norm Parseval frame} \cite{Casazza2004,CL} or 
a {\em finite normalized tight frame} \cite{Bene}. 
\subsection{Normalized Parseval frames versus orthonormal basis and stochastic aspects}
\label{fraorba}

Let us view the $N$ components of the vector $|u_i\rangle$ with respect to  the orthonormal basis $\{ |j\rangle \}_{j=1}^N$ as the respective conjugates of  $N$ functions  $i\mapsto\phi_j(i) $:
\begin{equation}
\nonumber |u_i\rangle = \sum_{j=1}^N \bar \phi_j(i) |j\rg\, ,
\end{equation}
(``bar'' means complex conjugate).
By using this expansion in  the resolution of the unity (\ref{ff}) we find the following orthogonality relations 
\begin{equation}
\label{orthphi}
(\phi_j, \phi_k)_{\kappa} = \delta_{jk}\, ,
\end{equation} 
with respect to the scalar product defined on the $M$-dimensional vector space of real or complex valued functions $i\mapsto \phi(i) $  on the set $X = \{1,2, \ldots, M\}$ by:
\begin{equation}
\label{inprodkappa}
(\phi, \phi')_{\kappa} \deq \sum_{i=1}^M \kappa_i \bar\phi(i)\, \phi' (i)\, .
\end{equation} 
By introducing the $N\times M$ matrix $L$ with matrix elements
\begin{equation}
\nonumber L_{ji}= \sqrt{\kappa_i} \bar\phi_j(i) = \sqrt{\kappa_i} \lg j|u_i\rg\, ,
\end{equation}
we easily derive from (\ref{orthphi}) the equation
\begin{equation}
\label{LLdag}
L\,L^{\dag} = \mathbb{I}_{\mathcal{H}}\,.
\end{equation}
Let us now express the pair overlaps $\lg u_i | u_{i'}\rg$ in terms of the functions $\phi_j$:
\begin{equation}
\label{LdagL}
\lg u_i | u_{i'}\rg = \sum_{j=1}^N \phi_j (i) \bar\phi_j (i')= \left(\mathrm{K}^{-1/2}L^{\dag}L\mathrm{K}^{-1/2}\right)_{ii'}\, , 
\end{equation}
where $\mathrm{K} \deq \mathrm{diag}(\kappa_1,\kappa_2, \ldots, \kappa_M)$.  If $M= N$, then (\ref{LLdag}) implies $L^{\dag}= L^{-1}$ and so $\lg u_i | u_{i'}\rg = \delta_{ii'}/\kappa_i$. The latter orthogonality relations  together with (\ref{LdagL}) implies that $\kappa_i = 1$ for all $i$ since the vectors $|u_i\rg$'s are all unit. As expected, any family of $N$ vectors satisfying (\ref{ff}) is an orthonormal basis.  

Let us introduce the real $M\times M$ matrix $U$ with matrix elements
\begin{equation}
\label{matrixM}
U_{ij}= \vert \lg u_i | u_j \rg \vert^2\, .
\end{equation}
These elements obey $U_{ii}=1$ for $1\leq i \leq M$ and $0\leq U_{ij} =  U_{ji}\leq 1$ for any pair $(i,j)$, with $i\neq j$. 

Now we  suppose that there is no pair of orthogonal elements, i.e. $0<  U_{ij}$ if $i\neq j$,   and no pair of proportional elements, i.e.   $  U_{ij}< 1 $ if $i\neq j$, in the frame.  Then from the Perron-Frobenius theorem for (strictly) positive matrices, the rayon spectral $r=r(U)$ is $>0$ and is  dominant simple eigenvalue of $U$. There exists a unique vector, $v_r$, $\Vert v_r \Vert = 1$, which is strictly positive (all components are $>0$) and $U v_r = r v_r$. All other eigenvalues $\alpha$  of $U$ lie within the open disk of radius $r$  : $\vert \alpha \vert <  r$. Since $\mathrm{tr}\, U = M$, and that $U$ has $M$ eigenvalues, one should have $r > 1$. The value $r=1$ represents precisely the limit case in which all eigenvalues are $1$, i.e. $U= \mathbb{I}$ and the frame is just an orthonormal basis of $\mathbb{C}^M$. It is then natural to view the number 
\begin{equation}
\label{etapar}
\eta \deq r-1
\end{equation}
as a kind of ``distance'' of the frame to the orthonormality. The question is to find the relation between the set $\{\kappa_1, \kappa_2, ... ,\kappa_M\}$ of weights defining the frame and the distance $\eta$.  By projecting on each vector $|u_i\rg$ from both sides the frame resolution of the unity  (\ref{ff}), we easily obtain the $M$ equations
\begin{equation}
\label{projframe}
1 = \lg u_iÊ| u_i \rg = \sum_{j=1}^M \kappa_j \vert \lg u_i| u_j\rg \vert^2\, , \quad \mbox{i.e.}\quad Uv_{\kappa}= v_{\delta}\, , 
\end{equation}
where ${}^tv_{\kappa} \deq 
  ( \kappa_1 \,\kappa_2\, ...  \,\kappa_M  )$ and ${}^tv_{\delta} \deq 
   (1 \,1\, ... \,1 )$ is the first diagonal vector  in $\mathbb{C}^M$. In the ``uniform'' case for which $\kappa_i = N/M$ for all $i$, i.e. in the case of a finite equal norm Parseval frame, 
which means that $v_{\kappa} = (N/M)\, v_{\delta}$, then $r = M/N$ and $v_r = 1/\sqrt{M} \, v_{\delta}$. 
   In this case, the distance to orthonormality is just 
   \begin{equation}
\label{distortho1}
\eta = \frac{M-N}{N}\, ,
\end{equation}
a relation which clearly exemplifies what we can expect at the limit $N \to M$.

Another aspect of a frame is the (right) stochastic nature of the matrix $P \deq U\,K$, evident from (\ref{projframe}). The row vector $\varpi \deq {}^tv_{\kappa}/N = (\kappa_1/N\,  \kappa_2/N\, ...\, \kappa_M/N\, )$ is a stationary probability vector:
\begin{equation}
\label{probvect}
\varpi \, P = \varpi\, .
\end{equation}
As is well known, this vector obeys the ergodic property:
\begin{equation}
\label{ergprop}
\lim_{k \to \infty} \left(P^k\right)_{ij} = \varpi_j = \frac{\kappa_j}{N}\, .
\end{equation}
\subsection{Parseval frames obtained by projection}
Let $\mathcal{E}$ be a finite-dimensional Hilbert space over $\mathbb{K}$, and let 
$\{ |\varepsilon _1\rangle ,|\varepsilon _2\rangle ,...,|\varepsilon _M\rangle \}$ 
be an orthonormal basis in $\mathcal{E}$. A large class of tight frames can be obtained by projection \cite{Ch}.
\\[5mm]
{\bf Theorem 1}  {\it If $\{ |\phi _j\rangle \}_{j=1}^N$
is an orthonormal system in $\mathcal{E}$ then $\{ |w_i\rangle \}_{i=1}^M$, where 
\begin{equation}
|w_i\rangle \!=\!\sum_{j=1}^N|\phi _j\rangle \langle \phi _j|\varepsilon _i\rangle 
\end{equation}
is a Parseval frame in the subspace $\mathcal{H}={\rm span} \, \{ |\phi _1\rangle ,\, |\phi _2\rangle ,\, \dots \, ,\, |\phi _N\rangle \}$, that is,}
\[ 
\mathcal{H}=\sum_{j=1}^N\mathbb{K}\, |\phi _j\rangle 
=\left\{\ \left. \sum_{j=1}^N\alpha _j\, |\phi _j\rangle \ \right|\ \ \alpha _1,\, \alpha _2,\, \dots \, ,\, \alpha _N\in \mathbb{K} \ \right\} .
\]
{\it Proof.} We get
\[
\begin{array}{rl}
\sum_{i=1}^M |w_i\rangle \langle w_i| &
=\sum_{i=1}^M \left( \sum_{j=1}^N |\phi _j\rangle \langle \phi _j|\varepsilon _i\rangle \right)
\left( \sum_{k=1}^N \langle \varepsilon _i|\phi _k\rangle \langle \phi _k| \right)\\[5mm]
 & =\sum_{j,k=1}^N \left( \sum_{i=1}^M \langle \phi _j|\varepsilon _i\rangle 
\langle \varepsilon _i|\phi _k\rangle \right)|\phi _j\rangle \langle \phi _k|
=\mathbb{I}_\mathcal{H}.\qquad  \opensquare
\end{array}
\]
The operator $\pi \!=\!\sum _{j=1}^N|\phi _j\rangle \langle \phi _j|$ is the orthogonal
projector corresponding to $\mathcal{H}$ and 
$|w_i\rangle \!=\!\pi |\varepsilon _i\rangle $. 
If two orthonormal systems $\{ |\phi _1\rangle , |\phi _2\rangle ,..., |\phi _N\rangle \}$
and $\{ |\psi _1\rangle , |\psi _2\rangle ,..., |\psi _N\rangle \}$ span the same subspace 
$\mathcal{H}$ then they define the same frame in $\mathcal{H}$. This means that the frame
depends on the subspace $\mathcal{H}$ we choose, and not on the particular orthonormal system we use.
\subsection{Embedding into a superspace defined by a Parseval frame}
Let $\mathcal{H}$ be a Hilbert space over $\mathbb{K}$, $\{ |j\rangle \}_{j=1}^N$ 
an orthonormal basis in $\mathcal{H}$, and let 
$\{ |e_i\rangle \}_{i=1}^M$ be the canonical basis of $\mathbb{K}^M$. The following result, proved independently by Naimark and 
Han/Larson \cite{Ch,H} shows that any finite Parceval frame can be obtained by projection. \\[5mm] 
{\bf Theorem 2}  \ {\it  
a) If $ \{|w_i\rangle \}_{i=1}^M$ is a Parseval frame in $\mathcal{H}$ 
then the system $\{ |\phi _j\rangle \}_{j=1}^N$, where 
\begin{equation}
|\phi _j\rangle \!=\!\sum_{i=1}^M |e_i\rangle \langle w_i|j\rangle 
\!=\!(\langle w_1|j\rangle , \langle w_2|j\rangle ,...,\langle w_M|j\rangle )
\end{equation}
is an orthonormal system in $\mathbb{K}^M$.\\[3mm] 
b) The Hilbert space $\mathcal{H}$ can be identified with the subspace
\[ 
\tilde \mathcal{H}={\rm span} \, \{ |\phi _1\rangle ,\, |\phi _2\rangle ,\, \dots \, ,\, |\phi _N\rangle \}
\]
of the superspace $\mathbb{K}^M$ by using the isometry
$\mathcal{H}\longrightarrow \tilde \mathcal{H}:\ |v\rangle \mapsto |\tilde v\rangle $, where
\begin{equation}
|\tilde v\rangle =\sum_{j=1}^N|\phi _j\rangle \langle j|v\rangle 
=\sum_{i=1}^M|e_i\rangle \langle w_i|v\rangle \!=\!(\langle w_1|v\rangle , \langle w_2|v\rangle ,...,\langle w_M|v\rangle )
\end{equation}
c) The frame $ \{|\tilde w_i\rangle \}_{i=1}^M$ corresponding to $ \{|w_i\rangle \}_{i=1}^M$ is the orthogonal projection of 
the orthonormal basis  $\{ |e_i\rangle \}_{i=1}^M$ }
\begin{equation}
|\tilde w_i\rangle =\pi |e_i\rangle \qquad for\ any\quad i\in \{ 1,2,...,M\}.
\end{equation}
{\it Proof.} a) From (\ref{vscalarw}) we deduce that
$\langle \phi _j|\phi _k\rangle =\sum_{i=1}^M \langle j|w_i\rangle \langle w_i|k\rangle 
=\langle j|k\rangle =\delta _{jk}.$\\
b) We get 
$|\tilde v\rangle =\sum_{j=1}^N|\phi _j\rangle \langle j|v\rangle =
\sum_{j=1}^N \sum_{i=1}^M|e_i\rangle \langle w_i|j\rangle \langle j|v\rangle 
=\sum_{i=1}^M|e_i\rangle \langle w_i|v\rangle .$\\
c) We have $\pi |e_i\rangle =\sum_{j=1}^N|\phi _j\rangle \langle j|w_i\rangle =|\tilde w_i\rangle .\qquad  \opensquare $\\[5mm]
The subspace $\tilde \mathcal{H}$ and the isometry $\mathcal{H}\longrightarrow \tilde \mathcal{H}$ have been defined by using an 
orthonormal basis $\{ |j\rangle \}_{j=1}^N$ but they do not depend on the basis we choose.
The representation $|\tilde v\rangle $ of $|v\rangle $ can be regarded as a discrete counterpart  to the usual Fock-Bargmann representation \cite{G2000}.
\subsection{Finite tight frames defined by using groups}
Some useful frames can be defined in a natural way by using  group representations \cite{H}.
Let $\{ g\!:\!\mathcal{H}\longrightarrow \mathcal{H}\ |\ g\!\in \!G\ \}$ be an orthogonal (resp. unitary)
irreducible representation of a finite group $\mathcal{G}$ in the real (resp. complex) $n$-dimensional Hilbert space
$\mathcal{H}$, and let $|w\rangle \in \mathcal{H}$ be a fixed vector. The elements $g\in \mathcal{G}$ with the 
property
\begin{equation}
g|w\rangle =\alpha |w\rangle 
\end{equation}
where $\alpha $ is a scalar depending on $g$, form the stationary group $\mathcal{G}_w$ of $|w\rangle $.\\[5mm]
{\bf Theorem 3} {\it If $\{g_i\}_{i=1}^M$ is a system of representatives of the left cosets of $\mathcal{G}$ on $\mathcal{G}_w$
then 
\begin{equation}
|w_1\rangle =g_1|w\rangle , \quad |w_2\rangle =g_2|w\rangle ,\quad ...\quad |w_M\rangle =g_M|w\rangle 
\end{equation}
form an equal norm tight frame in $\mathcal{H}$, namely } 
\begin{equation} 
\sum_{i=1}^M |w_i\rangle \langle w_i|=\frac{M}{N}||w||^2\, \mathbb{I}_\mathcal{H}.
\end{equation}
{\bf Proof.} The operator $\Lambda :\mathcal{H}\longrightarrow \mathcal{H}$, 
$\Lambda |v\rangle =\sum_{i=1}^M|w_i\rangle \langle w_i|v\rangle $ is self-adjoint 
\[ \langle v'|(\Lambda |v\rangle )=\sum_{i=1}^M\langle v'|w_i\rangle \langle w_i|v\rangle =(\langle v'|\Lambda )|v\rangle \]
and therefore, it has a real eigenvalue $\lambda $. 
Since the eigenspace $\{ \, |v\rangle \ ; \ \Lambda |v\rangle =\lambda |v\rangle \, \}$ corresponding to $\lambda $ is $\mathcal{G}$-invariant
\[  \Lambda (g|v\rangle )\!=\!\sum_{i=1}^M|w_i\rangle \langle w_i|(g|v\rangle )
=\sum_{i=1}^Mg|w_i\rangle \langle w_i|v\rangle =g(\Lambda |v\rangle ) \]
and the representation  is irreducible we must have $\Lambda |v\rangle =\lambda |v\rangle $ for any $|v\rangle \!\in \!\mathcal{H}$.
By using an orthogonal basis $\{ |1\rangle ,|2\rangle ,...,|N\rangle \}$ of $\mathcal{H}$ we get
\[ N\lambda =\sum_{j=1}^N\langle j|\Lambda |j\rangle =\sum_{j=1}^N\sum_{i=1}^M\langle j|w_i\rangle \langle w_i|j\rangle 
=\sum_{i=1}^M \sum_{j=1}^N|\langle j|w_i\rangle |^2=M\, ||w||^2. \qquad  \opensquare \]
One can easily remark that the whole orbit
\[
G|w\rangle =\left\{ \, g|w\rangle \ |\ \ g\in G\, \right\}
\]
is a tight frame, and more than that, any finite union of orbits is also a tight frame. The relation
\begin{equation}\label{repCn}
g(\alpha _1,\alpha _2)=\left( \alpha _1\, \cos \frac{2\pi }{n}-\alpha _2\, \sin \frac{2\pi }{n},\, 
\alpha _1\, \sin \frac{2\pi }{n}+\alpha _2\, \cos \frac{2\pi }{n}\right)
\end{equation}
defines a representation of the cyclic group $\mathcal{C}_n=\langle \, g\, |\ g^n=e\, \rangle $ as a group of rotations of the plane,
and for example, the orbit
\begin{equation} \label{hl}
\begin{array}{l}
\mathcal{C}_3\left(\sqrt{\frac{2}{3}},0\right)=
\!=\!\left\{ \, \left(\sqrt{\frac{2}{3}},0\right),\, \left( -\frac{1}{\sqrt{6}},\frac{1}{\sqrt{2}}\right),\,   
\left( -\frac{1}{\sqrt{6}},-\frac{1}{\sqrt{2}}\right)\, \right\} 
\end{array}
\end{equation}
is a Parseval frame in $\mathbb{R}^2$. The relations
\begin{equation} \label{repT}
g(\alpha _1,\alpha _2,\alpha _3)=(-\alpha _1,-\alpha _2,\alpha _3),\qquad 
h(\alpha _1,\alpha _2,\alpha _3)=(\alpha _2,\alpha _3,\alpha _1)
\end{equation}
define a representation of the tetrahedral group $\mathcal{T}=\langle \, g,\,h\, |\ g^2=h^3=(gh)^3=e\, \rangle $ as a group of rotations of the space,
and for example,
\begin{equation} \label{ds}
\begin{array}{l}
T\left( -\frac{1}{2}, \frac{1}{2}, \frac{1}{2}\right)\!=\!
\left\{ \left( -\frac{1}{2}, \frac{1}{2}, \frac{1}{2}\right), 
\left( \frac{1}{2}, -\frac{1}{2}, \frac{1}{2}\right),
\left( \frac{1}{2}, \frac{1}{2}, -\frac{1}{2}\right), 
\left( -\frac{1}{2}, -\frac{1}{2}, -\frac{1}{2}\right) \right\}
\end{array}
\end{equation}
is a Parseval frame in $\mathbb{R}^3$.
%
%
%
%
\section{Integer coefficients}
Let $\mathcal{H}=\mathbb{R}^N$ and let $ \{|w_i \rangle \}_{i=1}^M$, where
\[
\begin{array}{l}
|w_1\rangle =(w_{11},w_{12},\dots ,w_{1N})\\
|w_2\rangle =(w_{21},w_{22},\dots ,w_{2N})\\
.....................................\\
|w_M\rangle =(w_{M1},w_{M2},\dots ,w_{MN})
\end{array}
\]
be a Parseval frame in $\mathbb{R}^N$, that is,
\[
\sum_{i=1}^M|w_i\rangle \langle w_i |v\rangle =|v\rangle \qquad {\rm for\ any\ } \ |v\rangle \in \mathbb{R}^N.
\]
In view of theorem 2 the vectors
\[
\begin{array}{l}
|\phi _1\rangle =(w_{11},w_{21},\dots ,w_{M1})\\
|\phi _2\rangle =(w_{12},w_{22},\dots ,w_{M2})\\
.....................................\\
|\phi _N\rangle =(w_{1N},w_{2N},\dots ,w_{MN})
\end{array}
\]
form an orthonormal system in $\mathcal{E}=\mathbb{R}^M$, and the injective mapping (analysis operator)
\[
T:\mathbb{R}^N\longrightarrow \mathbb{R}^M:|v\rangle \mapsto |\tilde v\rangle 
\!=\!(\langle w_1|v\rangle , \langle w_2|v\rangle ,...,\langle w_M|v\rangle )
\]
 which can be written as
\[
\mathbb{R}^N\longrightarrow \mathbb{R}^M,\qquad  
T(\alpha _1,\alpha _2,\dots ,\alpha _N)= \alpha _1\, |\phi _1\rangle +\alpha _2\, |\phi _2\rangle +\cdots \alpha _N\, |\phi _N\rangle 
\]
allows us to identify $\mathbb{R}^N$ with the subspace
\[
\tilde \mathcal{H}=\{\ \alpha _1|\phi _1\rangle +\alpha _2|\phi _2\rangle +...+\alpha _N|\phi _N\rangle \ |
\ \ \alpha _1, \, \alpha _2,\, ...,\ \alpha _N \in \mathbb{R} \ \} 
\]
of the superspace $\mathbb{R}^M$. The one-to-one mapping $\mathbb{R}^N\longrightarrow \tilde \mathcal{H}:|v\rangle \mapsto |\tilde v\rangle $
is an isometry 
\[ 
\langle \tilde v|\tilde v'\rangle =\langle v|v'\rangle ,\qquad \qquad ||\tilde v||=||v||
\]
and $ \{|\tilde w_i \rangle \}_{i=1}^M$ is a Parseval frame in $\tilde \mathcal{H}$ corresponding to $ \{|w_i \rangle \}_{i=1}^M$
\[
\sum_{i=1}^M|\tilde w_i\rangle \langle \tilde w_i |\tilde v\rangle =\sum_{i=1}^M|\tilde w_i\rangle \langle w_i |v\rangle =
\sum_{i=1}^MT|w_i\rangle \langle w_i |v\rangle =T|v\rangle =|\tilde v\rangle .
\]
The  frame $ \{|\tilde w_i \rangle \}_{i=1}^M$ is the orthogonal projection on $\tilde \mathcal{H}$ of the canonical basis
\[
|e_1\rangle =(1,0,0\dots ,0),\quad |e_2\rangle =(0,1,0,\dots ,0),\quad \dots \quad |e_M\rangle =(0,0,\dots ,0,1)
\]
namely, by denoting $\pi \!=\! \sum_{j=1}^N|\phi _j\rangle \langle \phi _j|$, we have
\[
|\tilde w_1\rangle =\pi |e_1\rangle ,\qquad |\tilde w_2\rangle =\pi |e_2\rangle ,\qquad \dots\quad |\tilde w_M\rangle =\pi |e_M\rangle .
\]
The matrix of $\pi $ in terms of the canonical basis $\{ |e_i\rangle \}_{i=1}^M$  is 
\begin{equation}
\pi =\left( \begin{array}{cccc}
\langle w_1|w_1\rangle  & \langle w_1|w_2\rangle & ... & \langle w_1|w_M\rangle \\
\langle w_2|w_1\rangle & \langle w_2|w_2\rangle   & ... & \langle w_2|w_M\rangle \\
... & ... & ... & ...\\
\langle w_M|w_1\rangle & \langle w_M|w_2\rangle  & ...  & \langle w_M|w_M\rangle 
\end{array} \right) 
\end{equation}
The linear operator 
\[
\pi ^\perp :\mathbb{R}^M\longrightarrow \mathbb{R}^M,\qquad \pi ^\perp x=x-\pi x 
\]
is the orthogonal projector corresponding to the orthogonal complement 
\begin{equation}
\tilde \mathcal{H}^\perp =\left\{ x=(x_1,x_2,...,x_M)\ \left| \  
\sum_{i=1}^Mx_i\, |w_i\rangle =0 \right. \right\}.
\end{equation}
of $\tilde \mathcal{H}$ in $\mathbb{R}^M$, and the vectors 
\[
|\tilde w_1^\perp \rangle =\pi^\perp  |e_1\rangle ,\qquad 
|\tilde w_2^\perp \rangle =\pi ^\perp |e_2\rangle ,\qquad \dots\quad 
|\tilde w_M^\perp \rangle =\pi |e_M^\perp \rangle .
\]
form a frame $ \{|\tilde w_i ^\perp \rangle \}_{i=1}^M$ in $\tilde \mathcal{H}^\perp $ such that
\[
|\tilde w_i\rangle +|\tilde w_i^\perp \rangle =|e_i\rangle \qquad {\rm for\ any}\quad i\in \{ 1,2,...,M\}
\]
called  the {\it complementary frame} \cite{H}. Particularly, one can remark that the complementary frame corresponding 
to an equal norm frame is an equal norm frame.\\[5mm]
Each vector $|v\rangle \in \mathbb{R}^N$ can be written as a linear combination of the frame vectors $|w_i\rangle $
\[
|v\rangle =\sum_{i=1}^N|w_i\rangle \langle w_i|v\rangle 
\]
in terms of the {\em frame coefficients} $\langle w_i|v\rangle $.
If $M>N$ then the representation of a vector $|v\rangle \in \mathcal{H}$ as a linear combination of 
the frame vectors is not unique, and we have 
\[
|v\rangle =\sum_{i=1}^Nx_i\, |w_i\rangle 
\]
that is, the relation 
\[
\sum_{i=1}^Nx_i\, |w_i\rangle =\sum_{i=1}^N|w_i\rangle \langle w_i|v\rangle 
\]
which can be written as
\[
\sum_{i=1}^N(x_i-\langle w_i|v\rangle )\, |w_i\rangle =0
\]
if and only if 
\[
(x_1-\langle w_1|v\rangle ,\,x_2-\langle w_2|v\rangle ,\, \dots\, ,x_M-\langle w_M|v\rangle )\in \tilde \mathcal{H}^\perp 
\]
that is, if and only if
\[
(x_1,\,x_2,\, \dots\, ,x_M)\in 
(\langle w_1|v\rangle ,\,\langle w_2|v\rangle ,\, \dots\, ,\langle w_M|v\rangle )+\tilde \mathcal{H}^\perp .
\]
From the last relation it follows
\[
|v\rangle =\sum_{i=1}^Nx_i\, |w_i\rangle \ \ \Longleftrightarrow \ \ \pi (x_1,x_2,...,x_M)=
(\langle w_1|v\rangle ,\,\langle w_2|v\rangle ,\, \dots\, ,\langle w_M|v\rangle )
\]
and the inequality obtained by Duffin and Schaeffer \cite{Duffin}
\begin{equation}
|v\rangle =\sum_{i=1}^Mx_i\, |w_i\rangle \qquad \Longrightarrow \qquad \sum_{i=1}^M(x_i)^2\geq \sum_{i=1}^M(\langle w_i|v\rangle )^2.
\end{equation}
Each vector $|v\rangle \in \mathbb{R}^N$ admits a natural representation in terms of frame coefficients $\langle w_i|v\rangle $, but other
representations may offer additional facilities. In certain applications it is advantageous \cite{CDOSZ} to replace the frame coefficients
by {\em quantized coefficients}, i.e. by integer multiples of a given $\delta >0$. In this section we shall present some applications 
concerning the elements of a Hilbert space which can be written as a linear combination  with integer coefficients of the vectors of a fixed frame.
\subsection{Orthogonal projection of \ $\mathbb{Z}^M$ on a subspace of \ $\mathbb{R}^M$}
Let $E$ be a vector subspace of $\mathbb{R}^M$ and let $B_r(a)=\{ \, x\in E\ |\ \|x-a\|<r\, \}$ be the open ball of center $a$ and radius $r$.
A set $D\subset E$ is {\em dense} in $E$ if the ball $B_r(a)$ contains at least a point of $D$ for any $a\in E$ and any $r\in (0,\infty )$.
The set $D$ is {\em relatively dense} in $E$ if there is $r\in (0,\infty )$ such that the ball $B_r(a)$ contains at least a point of $D$ for 
any $a\in E$. The set $D$ is {\em discrete} in $E$ if for each $a\in D$ there is  $r\in (0,\infty )$ such that $D\cap B_r(a)=\{a\}$.
The set $D$ is {\em uniformly discrete} in $E$ if there is $r\in (0,\infty )$ such that the ball $B_r(a)$ contains at most one point of $D$ for 
any $a\in E$. The set $D$ is a {\em Delone set} in $E$ if it is both relatively dense and uniformly discrete in $E$. The set $D$ is a {\em lattice}
in $E$ if it is both an additive subgroup of $E$ and a Delone set in $E$. In order to describe the orthogonal projection of $\mathbb{Z}^M$ on $E$
we will use the following result.\\[5mm]
{\bf Theorem 4} \cite{Des,S} {\it Let $\Phi :\mathbb{R}^M\longrightarrow \mathbb{R}^L$ be a surjective linear mapping, where $L<M$.
Then there are subspaces $V$, $V'$ of $\mathbb{R}^L$ such that
\begin{itemize}
\item[a)] $\mathbb{R}^L=V\oplus V'$
\item[b)] $\Phi (\mathbb{Z}^M)=\Phi (\mathbb{Z}^M)\cap V+\Phi (\mathbb{Z}^M)\cap V'$
\item[c)] $\Phi (\mathbb{Z}^M)\cap V'$ is a lattice in $V'$
\item[d)] $\Phi (\mathbb{Z}^M)\cap V$ is a dense subgroup of $V$.
\end{itemize}
The subspace $V$ in this decomposition is uniquely determined.}\\[5mm]
The theorem 4 allows us to describe the subsets
\[
\pi (\mathbb{Z}^M)\!=\!\sum_{i=1}^M\mathbb{Z}\, |\tilde w_i\rangle 
\!=\!\left\{ \left. \, \sum_{i=1}^Mn_i\, |\tilde w_i\rangle \ \right| \ \ n_1,n_2,\dots ,n_M\!\in \!\mathbb{Z}\ \right\}
\]
of $\tilde \mathcal{H}$ and 
\[
\pi ^\perp (\mathbb{Z}^M)\!=\!\sum_{i=1}^M\mathbb{Z}\, |\tilde w_i^\perp \rangle 
\!=\!\left\{ \left. \, \sum_{i=1}^Mn_i\, |\tilde w_i^\perp \rangle \ \right| \ \ n_1,n_2,\dots ,n_M\!\in \!\mathbb{Z}\ \right\}
\]
of $\tilde \mathcal{H}^\perp $. There are subspaces $V$, $V'$ of $\tilde \mathcal{H}$ and subspaces $W$, $W'$ of $\tilde \mathcal{H}^\perp $
such that
\begin{equation}
\begin{array}{cc}
\tilde \mathcal{H}=V\oplus V' \qquad & \pi (\mathbb{Z}^M)=\pi (\mathbb{Z}^M)\cap V+\pi (\mathbb{Z}^M)\cap V'\\[2mm]
\tilde \mathcal{H}^\perp =W\oplus W'\qquad  & \pi ^\perp (\mathbb{Z}^M)\!=\!\pi ^\perp (\mathbb{Z}^M)\cap W\!+\!\pi ^\perp (\mathbb{Z}^M)\cap W'
\end{array}
\end{equation}
$\pi (\mathbb{Z}^M)\cap V'$ is a lattice in $V'$, \ $\pi ^\perp (\mathbb{Z}^M)\cap W'$ is a lattice in $W'$, \ 
$\pi (\mathbb{Z}^M)\cap V$ is a dense subgroup of $V$ and $\pi ^\perp  (\mathbb{Z}^M)\cap W$ is a dense subgroup of $W$.\\[5mm]
We say that the starting frame $ \{|w_i \rangle \}_{i=1}^M$ is a {\it periodic frame} if $V=\{ 0\}$, that is, if
\[
\sum_{i=1}^M\mathbb{Z}\, |w_i\rangle =\left\{ \left. \, \sum_{i=1}^Mn_i\, |w_i\rangle \ \right| \ \ n_1,n_2,\dots ,n_M\in \mathbb{Z}\ \right\}
\]
is a lattice in $\mathcal{H}$. The frame $ \{|w_i \rangle \}_{i=1}^M$ will be called a {\it quasiperiodic frame} if $W'=\{ 0\}$ and
$\pi $ restricted to $\mathbb{Z}^M$ is one-to-one. In this case, the collection of spaces and mappings
\begin{equation}
\begin{array}{cclcc}
\tilde \mathcal{H} & \stackrel{\pi }{\longleftarrow } & \mathbb{R}^M & \stackrel{\pi ^\perp}{\longrightarrow } & \tilde \mathcal{H}^\perp \\
 & & \cup & & \\
& & \mathbb{Z}^M & & 
\end{array}
\end{equation}
is a so-called {\em cut and project scheme} \cite{moodyjpg} and we can define the $\ast $-mapping
\begin{equation}
\pi (\mathbb{Z}^M)\longrightarrow \tilde \mathcal{H}^\perp :\ x\mapsto x^\ast =\pi ^\perp ((\pi |_{\mathbb{Z}^M})^{-1}x). 
\end{equation}
The projection $\pi $ restricted to $\mathbb{Z}^M$ is one-to-one if and only if $\mathbb{Z}^M\cap \tilde \mathcal{H}^\perp =\{ 0\}$.
The translations of $\tilde \mathcal{H}$ corresponding to the elements of $\mathbb{Z}^M\cap \tilde \mathcal{H}$ leave the set
$\pi ^\perp (\mathbb{Z}^M)$ invariant. If $\mathbb{Z}^M\cap \tilde \mathcal{H}$ contains a basis of $\tilde \mathcal{H}$ then the starting
frame is a periodic frame.
\subsection{Honeycomb lattice and diamond structure described in terms of frames}
The symmetry properties of certain discrete sets can be simpler described by using a frame instead of a basis.
Honeycomb lattice (figure 1) is a discrete subset $\mathcal{L}$ of the plane such that each point
$P\in \mathcal{L}$ has three nearest neighbours forming an equilateral triangle centered at $P$.
It can be described in a natural way by using the periodic Parseval frame (see (\ref{hl}))
\[ 
\begin{array}{l}
|w_1\rangle \!=\!\left(\sqrt{\frac{2}{3}},0\right),\qquad   |w_2\rangle \!=\!\left( -\frac{1}{\sqrt{6}},\frac{1}{\sqrt{2}}\right),\qquad  
|w_3\rangle \!=\!\left( -\frac{1}{\sqrt{6}},-\frac{1}{\sqrt{2}}\right)
\end{array}
\]
as the set \cite{C2006}
\[
\mathcal{L}=\{ \ n_1|w_1\rangle +n_2|w_2\rangle +n_3|w_3\rangle \ \ |\ \ (n_1,n_2,n_3)\in \mathbb{L}\ \}
\]
where the subset
\[
\mathbb{L}=\{ \ n=(n_1,n_2,n_3)\in \mathbb{Z}^3\ \ |\ \ n_1+n_2+n_3\in \{0,1\}\ \}
\]
of $\mathbb{Z}^3$ can be regarded as a mathematical model. The nearest neighbours of $n\in \mathbb{L}$ are
\[ 
\begin{array}{l}
n^1=(n_1+\nu (n),n_2,n_3)\\
n^2=(n_1,n_2+\nu (n),n_3)\\
n^3=(n_1,n_2,n_3+\nu (n))
\end{array} \qquad 
{\rm where}\quad \nu (n)=(-1)^{n_1+n_2+n_3}.
\]
The six points $n^{ij}=(n^i)^j$ corresponding to $i\not=j$ are the next-to-nearest neighbours,
and one can remark that $n^{ii}=n$, $n^{ijl}=n^{lji}$, for any $i,\, j,\, l\in \{ 1,2,3 \}$. The mapping 
\[ 
d:\mathbb{L}\times \mathbb{L}\longrightarrow \mathbb{Z}\qquad 
d(n,n')=|n_1-n'_1|+|n_2-n'_2|+|n_3-n'_3|
\]
is a distance on $\mathbb{L}$, and a point $n'$ is a neighbour of order $l$
of $n$ if $d(n,n')=l$.\\[5mm]
The symmetry group $\mathcal{G}$ of the honeycomb lattice is isomorphic with the group of
all the isometries of the metric space $(\mathbb{L},d)$, group generated by
the transformations 
\[
\begin{array}{l}
\mathbb{L}\longrightarrow \mathbb{L}:\ (n_1,n_2,n_3)\mapsto (n_2,n_3,n_1)\\
\mathbb{L}\longrightarrow \mathbb{L}:\ (n_1,n_2,n_3)\mapsto (n_1,n_3,n_2)\\
\mathbb{L}\longrightarrow \mathbb{L}:\ (n_1,n_2,n_3)\mapsto (-n_1\!+\!1,-n_2,-n_3).
\end{array}
\]
Honeycomb lattice is a mathematical model for a graphene sheet and the use of the indicated frame leads
to a simpler and more symmetric form for the $\mathcal{G}$-invariant mathematical objects occuring in the description
of certain physical properties \cite{C2006}.

\begin{figure}[h]
\setlength{\unitlength}{1mm}
\begin{picture}(110,50)(-8,0)
\multiput(18,0)(18,0){5}{\line(1,0){6}} %
\multiput(18,0)(18,0){6}{\line(-2,3){3}}
\multiput(9,4.5)(18,0){6}{\line(-2,-3){3}}
\multiput(9,4.5)(18,0){6}{\line(1,0){6}}
\multiput(9,4.5)(18,0){6}{\line(-2,3){3}}
\multiput(18,9)(18,0){6}{\line(-2,-3){3}}
\multiput(18,9)(18,0){5}{\line(1,0){6}} %
\multiput(18,9)(18,0){6}{\line(-2,3){3}}
\multiput(9,13.5)(18,0){6}{\line(-2,-3){3}}
\multiput(9,13.5)(18,0){6}{\line(1,0){6}}
\multiput(9,13.5)(18,0){6}{\line(-2,3){3}}
\multiput(18,18)(18,0){6}{\line(-2,-3){3}}
\multiput(18,18)(18,0){5}{\line(1,0){6}}%
\multiput(18,18)(18,0){6}{\line(-2,3){3}}
\multiput(9,22.5)(18,0){6}{\line(-2,-3){3}}
\multiput(9,22.5)(18,0){6}{\line(1,0){6}}
\multiput(9,22.5)(18,0){6}{\line(-2,3){3}}
\multiput(18,27)(18,0){6}{\line(-2,-3){3}}
\multiput(18,27)(18,0){5}{\line(1,0){6}}%
\multiput(18,27)(18,0){6}{\line(-2,3){3}}
\multiput(9,31.5)(18,0){6}{\line(-2,-3){3}}
\multiput(9,31.5)(18,0){6}{\line(1,0){6}}
\multiput(9,31.5)(18,0){6}{\line(-2,3){3}}
\multiput(18,36)(18,0){6}{\line(-2,-3){3}}
\multiput(18,36)(18,0){5}{\line(1,0){6}}%
\multiput(18,36)(18,0){6}{\line(-2,3){3}}
\multiput(9,40.5)(18,0){6}{\line(-2,-3){3}}
\multiput(9,40.5)(18,0){6}{\line(1,0){6}}
\multiput(9,40.5)(18,0){6}{\line(-2,3){3}}
\multiput(18,45)(18,0){6}{\line(-2,-3){3}}
\multiput(18,45)(18,0){5}{\line(1,0){6}}
\multiput(18,45)(18,0){6}{\line(-2,3){3}}
\multiput(9,49.5)(18,0){6}{\line(-2,-3){3}}
\multiput(9,49.5)(18,0){6}{\line(1,0){6}}
\thinlines
\put(52.8,12.7){{\scriptsize $|w_3\rangle $}}
\put(52.8,21.4){{\scriptsize $|w_2\rangle $}}
\put(61.2,17){{\scriptsize $|w_1\rangle $}}
\put(54,18){\circle*{1}}
\thicklines
\put(54,18){\vector(1,0){6}}
\put(54,18){\vector(-2,3){3}}
\put(54,18){\vector(-2,-3){3}}
\end{picture}
\caption{A fragment of the honeycomb lattice}
\end{figure}
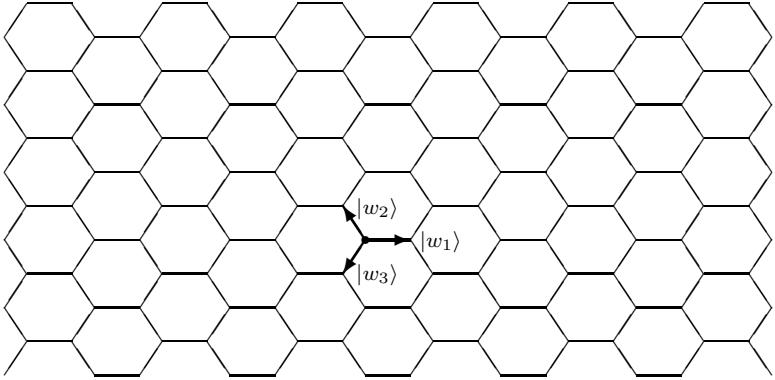

Diamond structure can be regarded as the three-dimensional analogue of the honeycomb lattice. Each point $P$
belonging to the diamond structure $\mathcal{D}$ has four nearest neighbours forming a regular tetrahedron centered at $P$. 
Diamond structure  can be described in a natural way by using the periodic Parseval frame (see (\ref{ds}))
\[
\begin{array}{ll}
|w_1\rangle =\left( -\frac{1}{2}, \frac{1}{2}, \frac{1}{2}\right),\qquad &
|w_2\rangle =\left( \frac{1}{2}, -\frac{1}{2}, \frac{1}{2}\right), \\[4mm]
|w_3\rangle =\left( \frac{1}{2}, \frac{1}{2}, -\frac{1}{2}\right), & 
|w_4\rangle =\left( -\frac{1}{2}, -\frac{1}{2}, -\frac{1}{2}\right) 
\end{array} 
\]
of $\mathbb{R}^3$ as the set \cite{C1995}
\[
\mathcal{D}=\{ \ n_1|w_1\rangle +n_2|w_2\rangle +n_3|w_3\rangle +n_4|w_4\rangle\ \ |\ \ 
(n_1,n_2,n_3,n_4)\in \mathbb{D}\ \}
\]
where 
\[
\mathbb{D}=\{ \ n=(n_1,n_2,n_3,n_4)\in \mathbb{Z}^4\ \ |\ \ n_1+n_2+n_3+n_4\in \{0,1\}\ \}.
\]
The nearest neighbours of a point $n\in \mathbb{D}$ are
\[
\begin{array}{l}
n^1=(n_1+\nu (n),n_2,n_3,n_4)\\
n^2=(n_1,n_2+\nu (n),n_3,n_4)\\
n^3=(n_1,n_2,n_3+\nu (n),n_4)\\
n^4=(n_1,n_2,n_3,n_4+\nu (n))
\end{array} \qquad 
{\rm where}\quad \nu (n)=(-1)^{n_1+n_2+n_3+n_4}.
\]
The twelve points $n^{ij}=(n^i)^j$ corresponding to $i\not=j$ are the next-to-nearest neighbours,
and one can remark that $n^{ii}=n$, $n^{ijl}=n^{lji}$, for any $i,\, j,\, l\in \{ 1,2,3,4 \}$. The mapping 
\[ 
d\!:\!\mathbb{D}\!\times \!\mathbb{D}\!\longrightarrow \!\mathbb{Z}\quad 
d(n,n')\!=\!|n_1\!-\!n'_1|\!+\!|n_2\!-\!n'_2|\!+\!|n_3\!-\!n'_3|\!+\!|n_4\!-\!n'_4|
\]
is a distance on $\mathbb{D}$, and a point $n'$ is a neighbour of order $l$
of $n$ if $d(n,n')=l$.\\[5mm]
The symmetry group $O_h^7$ of the diamond structure is isomorphic with the group of
all the isometries of the metric space $(\mathbb{D},d)$, group generated by
the transformations 
\[
\begin{array}{l}
\mathbb{D}\longrightarrow \mathbb{D}:\ (n_1,n_2,n_3,n_4)\mapsto (n_3,n_4,n_2,n_1)\\
\mathbb{D}\longrightarrow \mathbb{D}:\ (n_1,n_2,n_3,n_4)\mapsto (n_4,n_2,n_3,n_1)\\
\mathbb{D}\longrightarrow \mathbb{D}:\ (n_1,n_2,n_3,n_4)\mapsto (-n_1\!+\!1,-n_2,-n_3,-n_4).
\end{array}
\]
Again the use of a frame leads to a simpler and more symmetric form for the 
$O_h^7$-invariant mathematical objects occuring in the description of certain physical properties \cite{C1995}.
\subsection{An application to quasicrystals}
The group $\mathcal{I}$ of all the rotations of $\mathbb{R}^3$ leaving a regular icosahedron centered at the origin invariant 
is called the {\it icosahedral group}. The tvelwe points
\[
\pm (1,\tau,0), \ \pm (-1,\tau ,0),\ \pm (-\tau ,0,1),\ \pm (0,-1,\tau ),\ \pm(\tau ,0,1), \ \pm (0,1,\tau )  
\]
where $\tau =(1+\sqrt{5})/2$, are the vertices of a regular icosahedron centered at origin. The rotations 
\begin{equation} \label{repI} 
\begin{array}{l}
r(\alpha ,\beta  ,\gamma )\!=\!
\left(\frac{\tau \!-\!1}{2}\alpha \!-\!\frac{\tau }{2}\beta  \!+\!\frac{1}{2}\gamma ,
\, \frac{\tau }{2}\alpha \!+\!\frac{1}{2}\beta  \!+\!\frac{\tau -1}{2}\gamma ,
\, -\!\frac{1}{2}\alpha \!+\!\frac{\tau -1}{2}\beta  
\!+\!\frac{\tau }{2}\gamma \right)\\[1mm]
s(\alpha ,\beta  ,\gamma )=(-\alpha ,-\beta  ,\gamma ). 
\end{array} 
\end{equation}
satisfying the relation $r^5\!=\!s^2\!=\!(rs)^3\!=\!\mathbb{I}_{\mathbb{R}^3}$ leave this regular icosahedron invariant, and therefore
they define a representation of the icosahedral group in $\mathbb{R}^3$.

The stationary group $\mathcal{I}_{w}$ of $|w\rangle =\frac{1}{\sqrt{2(\tau +2)}}\, (1,\tau,0)$ is formed by the rotations $g\in \mathcal{I}$ 
with $g|w\rangle \in \{ |w\rangle ,-|w\rangle \}$, and we can choose the representatives $g_1$, $g_2$, ..., $g_6$ 
of the cosets of $\mathcal{I}$ on $\mathcal{I}_{w}$ such that 
\[
 \begin{array}{ll}
|w_1\rangle = g_1|w\rangle =\frac{1}{\sqrt{2(\tau +2)}}\, (1,\tau,0), & 
|w_2\rangle = g_2|w\rangle =\frac{1}{\sqrt{2(\tau +2)}}(-1,\tau ,0),\\[2mm]
|w_3\rangle = g_3|w\rangle =\frac{1}{\sqrt{2(\tau +2)}}\, (-\tau ,0,1),\quad  & 
|w_4\rangle = g_4|w\rangle =\frac{1}{\sqrt{2(\tau +2)}}\, (0,-1,\tau ),\\[2mm]  
|w_5\rangle = g_5|w\rangle =\frac{1}{\sqrt{2(\tau +2)}}\, (\tau ,0,1), & 
|w_6\rangle = g_6|w\rangle =\frac{1}{\sqrt{2(\tau +2)}}\, (0,1,\tau ).  
\end{array} 
\]
In view of theorem 3 the system $\{ |w_i\rangle \}_{i=1}^6$ is a tight frame in $\mathbb{R}^3$. By direct computation 
one can prove that it is a quasiperiodic Parseval frame
\[
\sum_{i=1}^6|w_i\rangle \langle w_i|=\mathbb{I}_{\mathbb{R}^3}.
\]
It defines an embedding of $\mathcal{H}=\mathbb{R}^3$ in the superspace $\mathbb{R}^6$ and the set
\begin{equation}
\mathcal{Q}=\{ \ x\in \pi (\mathbb{Z}^6)\ |\ x^*\in \pi ^\perp ([0,1]^6)\ \}
\end{equation}
defined by using the corresponding $*$-mapping is a quasiperiodic set.

The diffraction pattern corresponding to $\mathcal{Q}$ computed by using the Fourier transform 
is similar to the experimental diffraction patterns obtained in the case of certain icosahedral quasicrystals \cite{E,K}. 
Quasiperiodic sets corresponding to other quasicrystals can be obtained by starting from finite frames, 
and they help us to better understand the atomic structure of these materials.
\subsection{Sequences of finite frames}
Let $(f_n)_{n=0}^\infty $ be the Fibonacci sequence defined by reccurrence as 
\[
f_0=f_1=1,\qquad f_{n+1}=f_{n-1}+f_n
\]
and let $\tau _n=f_{n+1}/f_n$. It is well-known that $\lim_{n\rightarrow \infty }\tau _n=\tau $.
The tetrahedral frame $\mathcal{T}(1,\tau ,0)$ defined by using the representation (\ref{repT}) coincides with the icosahedral frame
$\mathcal{I}(1,\tau ,0)$ defined by using the representation (\ref{repI}) 
\[  
\begin{array}{l}
\mathcal{T}(1,\tau ,0)\!=\!\{ (1,\tau ,0), (-1,\tau ,0), (-\tau ,0,1), (0,-1,\tau ), \\[2mm]
 \qquad \qquad  \quad (\tau ,0,1), (0,1,\tau ), (-1,-\tau ,0), (1,-\tau ,0), \\[2mm]
 \qquad \qquad  \quad (\tau ,0,-1), (0,1,-\tau ), (-\tau ,0,-1), (0,-1,-\tau )\}\!=\!\mathcal{I}(1,\tau ,0).
\end{array}
\]
Therefore
\[ 
\begin{array}{l}
\lim_{n\rightarrow \infty }\mathcal{T}(1,\tau _n,0)=\lim_{n\rightarrow \infty }\{ (1,\tau _n,0), (-1,\tau _n,0), (-\tau _n,0,1), \\[2mm]
 \qquad \qquad (0,-1,\tau _n),(\tau _n,0,1), (0,1,\tau _n),
 (-1,-\tau _n,0), (1,-\tau _n,0), \\[2mm]
 \qquad \qquad (\tau _n,0,-1),(0,1,-\tau _n), (-\tau _n,0,-1), (0,-1,-\tau _n)\}=\mathcal{I}(1,\tau ,0)
\end{array}
\]
that is, we can approximate the frame $\mathcal{I}(1,\tau ,0)$ by using the periodic frames $\mathcal{T}(1,\tau _n,0)$.

The orbit $\mathcal{T}((1-t)(1,2,0)+t(1,\tau ,0))$
of the tetrahedral group $\mathcal{T}$ is a frame in $\mathbb{R}^3$ for any $t\in [0,1]$. It can be regarded as a continuous deformation
of the periodic frame $\mathcal{T}(1,2,0)$ into the icosahedral frame $\mathcal{I}(1,\tau ,0)$.

The relation
\begin{equation}
\mathcal{R}_\theta (x,y)=(x\, \cos \theta -y\, \sin \theta ,x\, \sin \theta +y\cos \theta )
\end{equation}
defines an $\mathbb{R}$-irreducible two-dimensional representation of the multiplicative group
\[
SO(2)=\left\{ \left. \left( \begin{array}{cc}
\cos \theta & -\sin \theta \\[2mm]
\sin \theta & \cos \theta 
\end{array} \right) \right| \ \ \theta \in [0,2\pi )\ \right\} 
\]
and the orbit $\{ \ |\theta \rangle =(\cos \theta ,\sin \theta )\ |\ \theta \in [0,2\pi )\ \}$ is a continuous frame 
\[
\frac{1}{\pi }\int_0^{2\pi }d\theta \, |\theta \rangle \langle \theta |=\mathbb{I}_{\mathbb{R}^2}.
\]
For any $n\in \mathbb{N}$ the orbit of $\mathcal{C}_n$ corresponding to $(1,0)$, namely,
\[
\mathcal{C}_n(1,0)=\left\{ \ \left. \left|\frac{2\pi }{n}k \right\rangle =\left( \cos \frac{2\pi }{n}k ,
\ \sin \frac{2\pi }{n}k \right)\ \right|\ k\in \{ 0,1,...,n-1\}\ \right\}
\]
is a finite frame
\[
\frac{2}{n}\sum_{k=0}^{n-1}\left|\frac{2\pi }{n}k \right\rangle  \left\langle \frac{2\pi }{n}k \right|
=\mathbb{I}_{\mathbb{R}^2}
\]
and we have
\[
\frac{2}{n}\sum_{k=0}^{n-1}\left|\frac{2\pi }{n}k \right\rangle  \left\langle \frac{2\pi }{n}k \right|
=\frac{1}{\pi }\, \frac{2\pi }{n}\sum_{k=0}^{n-1}\left|\frac{2\pi }{n}k \right\rangle  
\left\langle \frac{2\pi }{n}k \right|\stackrel{n\rightarrow \infty }{\longrightarrow}
\frac{1}{\pi }\int_0^{2\pi } d\theta \, |\theta \rangle \langle \theta |.
\]
Therefore, we can regard the continuous frame $\{ |\theta \rangle \}_{\theta \in [0,2\pi ]}$ as the limit of the sequence
of finite frames $(\, \mathcal{C}_n(1,0)\, )_{n=3}^\infty $.
%
%
\section{Frame quantization of discrete variable functions}
\subsection{Finite frame quantization}
\label{framequant}
Let $\mathcal{X}=\{ a_1,a_2,...,a_M\}$ be a fixed finite set we regard as a set of  data concerning a physical system. 
The space of all the functions 
$\varphi \!:\!\mathcal{X}\longrightarrow \mathbb{K}$ is a Hilbert space with the scalar product 
\begin{equation}
 \langle \varphi |\psi \rangle =\sum_{i=1}^M\overline{\varphi (a_i)}\, \psi (a_i)
\end{equation}
(evidently, if $\mathbb{K}=\mathbb{R}$
then $\overline{\varphi (a_i)}=\varphi (a_i)$) and the isometry
\begin{equation}
l^2(\mathcal{X})\longrightarrow \mathbb{K}^M:\ \varphi \mapsto (\varphi (a_1),\varphi (a_2),...,\varphi (a_M))
\end{equation}
allows us to identify the space $l^2(\mathcal{X})$ with the usual $M$-dimensional Hilbert space $\mathbb{K}^M$.
The system of functions $\{ \delta _1,\delta _2,...,\delta _M\}$, where
\[
\delta _i:\mathcal{X}\longrightarrow \mathbb{K}, \qquad \delta _i(a)=\left\{
\begin{array}{lll}
1 & {\rm if} & a=a_i\\
0 & {\rm if} & a\not=a_i
\end{array} \right. 
\]
is an orthonormal basis in $l^2(\mathcal{X})$
\[
\varphi =\sum_{i=1}^M\langle \delta _i|\varphi \rangle \delta _i=\sum_{i=1}^M\varphi (a_i)\delta _i.
\]
Let us select among the elements of $l^2(\mathcal{X})$ an orthonormal set 
$\{ \phi _1,\phi _2,...,\phi _N\}$ such that
\[
\kappa _i=\sum_{j=1}^N|\phi _j(a_i)|^2\not=0,\qquad {\rm for\ all\ } \ i\in \{ 1,2,...,M\}
\]
and let $\mathcal{H}={\rm span}\{ \phi _1,\phi _2,...,\phi _N\}$. 
In view of theorem 1, the elements 
\begin{equation}
|u_i\rangle \!=\!\frac{1}{\sqrt{\kappa _i}}\sum_{j=1}^N\langle \phi _j|\delta _i\rangle \phi _j
\!=\!\frac{1}{\sqrt{\kappa _i}}\sum_{j=1}^N\overline{\phi _j(a_i) }\,\phi _j,\qquad
i\!\in \!\{ 1,2,...,M\}
\end{equation}
form a normalized Parseval frame in $\mathcal{H}$, namely,
\[
\sum_{i=1}^M\kappa _i|u_i\rangle \langle u_i|=\mathbb{I}_\mathcal{H}.
\]

To each function $f:\mathcal{X}\longrightarrow \mathbb{R}$ which we regard as a {\it classical} observable
we associate the linear operator
\begin{equation}
A_f:\mathcal{H}\longrightarrow \mathcal{H},\qquad A_f=\sum_{i=1}^M\kappa _i\, f(a_i)\, |u_i\rangle \langle u_i|.
\end{equation}
This can be regarded as a Klauder-Berezin-Toeplitz type quantization \cite{B,K1963a,K1963b,K1995} of $f$, 
the notion of quantization being considered here in a 
wide sense \cite{G2007a,G2009,G2004a,G2004b, G2006, G2007b,G2003}. 
The eigenvalues of the matrix $A_f$ form the ``quantum spectrum'' of $f$  
(by opposition to its ``classical spectrum'' that is the set of its values $f(a_i)$). 
The function $f$ is called {\it upper (or contravariant) symbol} of $A_f$, and the function 
\begin{equation}
\check f:\ \mathcal{X}\longrightarrow \mathbb{R},\qquad  
\check f(a_k)=\langle u_k|A_f|u_k\rangle =\sum_{i=1}^M\kappa _i\, f(a_i)\, |\langle u_i|u_k\rangle |^2
\end{equation}
is called {\it lower (or covariant) symbol} of $A_f$. Since
\[
\sum_{k=1}^M\kappa _i\, |\langle u_i|u_k\rangle |^2=||u_i||^2=1
\]
the number $\check f(a_k)$ is a weighted mean of $f(a_1)$, $f(a_2)$, \dots $f(a_M)$, for any $k\in \{ 1,2,\dots M\}$.
In terms of the superspace, $\check f(a_k)$ can be regarded as a scalar product
\[
\check f(a_k)=\langle (f(a_1),\dots ,f(a_M)),
(\kappa _1\, |\langle u_1|u_k\rangle |^2,\dots ,\kappa _M\, |\langle u_M|u_k\rangle |^2)\rangle.
\]

To a certain extent, a quantization scheme consists in adopting a certain point of view in dealing with $\mathcal{X}$.
The presented frame quantization $f\mapsto A_f$ depends on the subspace $\mathcal{H}\subset l^2(\mathcal{X})$ we choose.
The validity of the frame quantization corresponding to a certain  subspace $\mathcal{H}$ is asserted by comparing 
spectral characteristics of $A_f$ with data provided by specific protocol in the observation of the considered 
physical system. An interesting subject of topological study is the triplet
$$
[M \ \mbox{values of}\ f]\leftrightarrow [N' \ \mbox{eigenvalues of}\ A_f\,, \,  N'\leq N] \leftrightarrow [M \ \mbox{values of}\ \check f]\,.
$$
\subsection{Probabilistic aspects of finite frame quantization}
The relations 
\begin{equation} 
\begin{array}{rlr}
\sum_{j=1}^N|\langle \phi _j|u_i\rangle |^2=1 & {\rm for} & i\in \{ 1,2,...,M\}\\[2mm] 
\sum_{i=1}^M\kappa _i\, |\langle \phi _j|u_i\rangle |^2=1 & {\rm for} & j\in \{ 1,2,...,N\}
\end{array} 
\end{equation}
show that the considered normalized Parseval frame defines two families of probability distributions.
This property can be interpreted in terms of a Bayesian duality \cite{algahel}.

If $\psi \in \mathcal{H}$ is such that 
$||\psi ||=\sqrt{\langle \psi ,\psi \rangle }=1$ then
\[
\sum_{i=1}^M|\sqrt{\kappa _i}\langle \psi |u_i\rangle |^2=
\sum_{i=1}^M\kappa _i\, |\langle \psi |u_i\rangle |^2=||\psi ||^2=1
\]
and hence, adopting the vocabulary of quantum measurement, $|\sqrt{\kappa _i}\langle \psi |u_i\rangle |^2$ can be viewed as the probability 
to find $\psi $ in the state $|u_i\rangle $.

The trace of the operator $A_f$ depends on the lower symbol
\[
\begin{array}{rl}
{\rm tr}\, A_f =\sum_{k=1}^N\langle \phi _k|A_f|\phi _k\rangle 
& = \sum_{k=1}^N\sum_{i=1}^M\kappa _i\langle \phi _k|u_i\rangle \langle u_i|A_f|\phi _k\rangle \\[5mm]
& =\sum_{i=1}^M\kappa _i \sum_{k=1}^N\langle u_i|A_f|\phi _k\rangle \langle \phi _k|u_i\rangle \\[5mm]
& =\sum_{i=1}^M\kappa _i\langle u_i|A_f|u_i\rangle =\sum_{i=1}^M\kappa _i\, \check f(a_i).
\end{array}
\]
An interesting problem in our finite frame quantization is to compare the starting function $f$ with the lower symbol $\check f$. 
With the  stochastic matrix notations of subsection \ref{fraorba}, the relation 
\[
\check f(a_k)=\langle u_k|A_f|u_k\rangle =\sum_{i=1}^M\kappa _i\, f(a_i)\, |\langle u_i|u_k\rangle |^2
\]
is rewritten as
\begin{equation}
\label{checkfmat}
\mathbf{ \check f} = P \mathbf{ f}\, ,
\end{equation}
with ${}^t\mathbf{ f} \deq \left(f(a_1)\, f(a_2) \,... \, f(a_M)\right)$ and ${}^t\mathbf{\check  f} \deq \left(\check f(a_1)\, \check f(a_2) \,... \, \check f(a_M)\right)$. This formula is interesting because it can be iterated:
\begin{equation}
\label{checkfmatit}
\mathbf{ \check f}^{[k]} = P^k\mathbf{ f}\, ,\quad \mathbf{ \check f}^{[k]}= P \mathbf{ \check f}^{[k-1]}\,, \quad \mathbf{ \check f}^{[1]}\equiv \mathbf{ \check f}\,, 
\end{equation}
and so we find from the  property (\ref{ergprop}) of $P$ that the ergodic limit (or ``long-term average'') of the iteration stabilizes to  the ``classical'' average of the  observable $f$ defined as:
\begin{equation}
\label{limerg}
\mathbf{ \check f}^{[\infty]}  = \lg f \rg_{\mathrm{cl}} \, v_{\delta}\, ,\ \lg f \rg_{\mathrm{cl}} \deq \sum_{i=1}^M \frac{\kappa_i}{N} f(a_i)\, .
\end{equation}

\subsection{The classical limit of  finite frame quantization}
\label{classlim}
We can evaluate the ``distance'' between the lower symbol and its classical counterpart through the  inequality:
\begin{equation}
\label{estnorm1}
\Vert \mathbf{ \check f} - \mathbf{ f}\Vert_{\infty}\deq \max_{1 \leq k \leq M}|f(a_k)\!-\!\check f(a_k)|\leq \Vert \mathbb{I} - P\Vert_{\infty} \, \Vert \mathbf{ f}\Vert_{\infty}\, ,  
\end{equation}
where the induced norm \cite{meyer} on matrix $A$ is $\Vert A\Vert_{\infty} = \max_{1 \leq i \leq M}\sum_{j=1}^M \vert a_{ij}\vert$.  In the present case, because of the stochastic nature of $P$, we have
\begin{equation}
\label{normP}
\Vert \mathbb{I} - P\Vert_{\infty} = 2\left(1 - \min_{1\leq i \leq M}\kappa_i\right)\, .
\end{equation}
In the uniform case, $\kappa_i = N/M$ for all $i$, we thus have an estimate of how far the two functions $f$ and $\check f$ are: $\Vert \mathbf{ \check f} - \mathbf{ f}\Vert_{\infty} \leq 2(M-N)/M\, \Vert \mathbf{ f}\Vert_{\infty}$. In the general case, we can view the parameter 
\begin{equation}
\label{zetapar}
\zeta \deq 1 - \min_{1\leq i \leq M}\kappa_i
\end{equation}
 as a distance of the ``quantum world''  to the classical one, of non-commutativity to commutativity, or again of the frame to orthonormal basis, like the distance $\eta = r-1$ introduced in subsection \ref{fraorba}.
 Another way to check that $\zeta = 1 - N/M \to 0$ means, in the uniform case $\kappa_i = N/M$ for all $i$, that  we go back to the classical spectrum of the observable $f$ results from the following relations. 
We have 
\[
\frac{N}{M}\sum_{j\not=k} |\langle u_j|u_k\rangle |^2=||u_k||^2-\frac{N}{M}|\langle u_k|u_k\rangle |^2=1-\frac{N}{M}
\]
and from the relation
\[
\check f(a_k)=\frac{N}{M}\sum_{j=1}^M\, f(a_j)\, |\langle u_j|u_k\rangle |^2 
=\frac{N}{M}\, f(a_k)+\sum_{j\not=k}\, f(a_j)\, |\langle u_j|u_k\rangle |^2
\]
we get
\[
\left( 1-\frac{N}{M}\right) \, \min_jf(a_j)\leq \check f(a_k)-\frac{N}{M}f(a_k)\leq \left( 1-\frac{N}{M}\right)\, \max_jf(a_j).
\]
 
Finally,  note  the estimates:
\[
|f(a_k)\!-\!\check f(a_k)|\!=\!\left| \sum_{i=1}^M\kappa _i(f(a_k)\!-\!f(a_i))\, |\langle u_i|u_k\rangle |^2 \right|
\!\leq \!\max_{i} |f(a_k)\!-\!f(a_i)|
\]
whence
\begin{equation}
\Vert \mathbf{ \check f} - \mathbf{ f}\Vert_{\infty} \leq \max_{i,k} |f(a_k)-f(a_i)|.
\end{equation}
\subsection{Frames defined by using eigenvectors of non-commuting operators}
Let $A,B:\mathbb{H}\longrightarrow \mathbb{H}$ be two operators on a Hilbert space $\mathbb{H}$, which are diagonalizable operators with orthogonal eigenvectors. 
If $AB=BA$ then there is a basis of $\mathbb{H}$ formed by common eigenvectors of $A$ and $B$, useful in the study of the 
operators which can be expressed as a function of $A$ and $B$. 
Such a basis does not exist if $AB\not=BA$, but a weaker version of this approach is possible by using a frame.
By starting from an orthonormal basis $\{ \varphi _i\}_{i=1}^M$ formed by eigenvectors of $A$ and an orthonormal basis $\{ \psi _j\}_{j=1}^M$
formed by eigenvectors of $B$ we can restrict us to a subspace of the form
\[
\mathcal{H}={\rm span}\{ \, \varphi _1,\, \varphi _2,\, \dots \, ,\, \varphi _N \}
\]
and use the frame $\{ \pi \psi _j\}_{j=1}^M$, where $\pi $ is the orthogonal projector corresponding to $\mathcal{H}$. 
In order to illustrate this method, let $\mathbb{Z}_M=\mathbb{Z}/M\mathbb{Z}=\{ 0,1,\dots ,M-1\}$ and
$A,B:l^2(\mathbb{Z}_M)\longrightarrow l^2(\mathbb{Z}_M)$ be the linear operators defined in terms of the canonical basis
$\{ \delta _i\}_{i=1}^M$ as
\[
A\delta _j=\delta _{j-1},\qquad B\delta _j={\rm e}^{\frac{2\pi {\rm i}}{M}j}\delta _j
\]
(the elements of $\mathbb{Z}_M$ are integers considered modulo $M$, and particularly, $-1=M-1$).

The elements of the canonical basis $\{ \delta _i\}_{i=1}^M$ are eigenfunctions of $B$. The functions
$\phi _0,\, \phi _1,\, ...\, \phi _{M-1}:\mathbb{Z}_M\longrightarrow \mathbb{C}$ defined as 
\begin{equation}
\phi _j(k)=\frac{1}{\sqrt{M}}{\rm e}^{-\frac{2\pi {\rm i}}{M}jk}
\end{equation}
that is,
\[
\phi _j=\frac{1}{\sqrt{M}}\sum_{k=0}^{M-1}{\rm e}^{-\frac{2\pi {\rm i}}{M}jk}\, \delta _k
\]
are eigenfunctions of $A$
\[
A\phi _j=\frac{1}{\sqrt{M}}\sum_{k=0}^{M-1}{\rm e}^{-\frac{2\pi {\rm i}}{M}jk}\, \delta _{k-1}={\rm e}^{-\frac{2\pi {\rm i}}{M}j}\, \phi _j
\]
and form an orthonormal basis in $l^2(\mathbb{Z}_M)$. 
Let $N\leq M$ and  $\mathcal{H}={\rm span}\{\phi _0,\, \phi _1,\, ...\, \phi _{N-1}\}$. The elements
\begin{equation}
|u_j\rangle =\frac{1}{\sqrt{N}}\sum_{k=0}^{N-1}{\rm e}^{\frac{2\pi {\rm i}}{M}jk}\phi _k \qquad 
j\in \{ 0,1, ...,M-1\}
\end{equation}
form a frame in the subspace $\mathcal{H}$ such that
\[
\frac{N}{M}\sum_{j=0}^{M-1} |u_j\rangle \langle u_j|=\mathbb{I}_\mathcal{H}.
\]
If $j\not= k$ then
\[
\langle u_j|u_k\rangle \!=\!\frac{1}{N}\sum_{p=0}^{N-1}{\rm e}^{\frac{2\pi {\rm i}}{M}(k-j)p}
\!=\!\frac{1}{N}\, \frac{1\!-\!{\rm e}^{\frac{2\pi {\rm i}}{M}(k-j)N}}{1\!-\!{\rm e}^{\frac{2\pi {\rm i}}{M}(k-j)}}
=\frac{{\rm e}^{\frac{\pi {\rm i}}{M}(k-j)(n-1)}}{N}\
\frac{\sin \frac{n\pi }{M}(k\!-\!j)} {\sin \frac{\pi }{M}(k\!-\!j)}.
\]
According to  the quantization scheme defined in subsection \ref{framequant},  the considered frame allows us to associate to each function 
$f:\mathbb{Z}_M\longrightarrow \mathbb{R}$ the operator 
\[
A_f:\mathcal{H}\longrightarrow \mathcal{H}, \quad 
A_f=\frac{N}{M}\sum_{k=0}^{M-1}f(k)\, |u_k\rangle \langle u_k|
\]
having the lower symbol 
\[
\begin{array}{l}
\check f(j)=\langle u_j|A_f|u_j\rangle =\frac{N}{M}\sum_{k=0}^{M-1}f(k)\, |\langle u_j|u_k\rangle |^2\\[5mm]
\qquad \qquad \qquad \quad \ \ =\frac{N}{M}f(j)+\frac{N}{M}\sum_{k\not=j}f(k)\, |\langle u_j|u_k\rangle |^2\\[5mm]
\qquad \qquad \qquad \quad \ \ =\frac{N}{M}f(j)+\frac{1}{NM}\sum_{k\not=j}f(k)\, 
\frac{\sin ^2\frac{n\pi }{M}(k-j)} {\sin ^2\frac{\pi }{M}(k-j)}.
\end{array}
\]
The entries of the matrix of $A_f$ in the orthonormal basis 
$\{ |\phi _0\rangle ,|\phi _1\rangle ,...,|\phi _{N-1}\rangle \}$ are
\begin{equation}
\langle \phi _p|A_f|\phi _q\rangle =\frac{1}{M}\sum_{k=0}^{M-1}{\rm e}^{\frac{2\pi {\rm i}}{M}k(p-q)}f(k)
\end{equation}
and
\begin{equation}
\begin{array}{r}
\langle u_j|A_f|u_j\rangle =\sum_{p,q=0}^{N-1}\langle u_j|\phi _p\rangle \, \langle \phi _p|A_f|\phi _q \rangle
\, \langle \phi _q|u_j\rangle \\[5mm]
=\frac{1}{N}\sum_{p,q=0}^{N-1}{\rm e}^{\frac{2\pi {\rm i}}{M}(q-p)j}\, \langle \phi _p|A_f|\phi _q \rangle .
\end{array}
\end{equation}
Particularly, we have
\[
\langle \phi _p|A_f|\phi _q\rangle =\frac{1}{M}\, \frac{1-a^M}{1-a{\rm e}^{\frac{2\pi {\rm i}}{M}(p-q)}}
\qquad {\rm in\ the\ case\ } \quad  f(k)=a^{k}
\]
and
\[
\langle \phi _p|A_f|\phi _q\rangle \!=\!\frac{1}{M}\left( 1\!+\!{\rm e}^{\frac{2\pi {\rm i}}{M}(p-q)}\right)^{M-1}
\quad {\rm in\ the\ case}\quad  f(k)\!=\!\!\left( \!\begin{array}{c}
M\!-\!1\\
k
\end{array}\!\right). 
\]
It is known that the functions $f_j:\mathbb{Z}_m\longrightarrow \mathbb{C}$ defined
in terms of Hermite polynomials
\[
f_j(k)=\sum_{l=-\infty }^\infty {\rm e}^{-\frac{\pi }{M}(lM+k)^2}\,  H_j\left( \sqrt{\frac{2\pi }{M}}(lM+k)\right)
\]
are eigenfunctions of the discrete Fourier transform \cite{M} 
\[
\frac{1}{\sqrt{M}}\sum_{p=0}^{M-1}{\rm e}^{\frac{2\pi {\rm i}}{M}pk}f_j(p)={\rm i}^j\, f_j(k).
\]
Therefore
\[
\langle \phi _p|A_{f_j}|\phi _q\rangle =\frac{{\rm i}^j}{\sqrt{M}}f_j(p-q).
\]

If the real number $x$ is not a multiple of $M$ then
\[
\sum_{k=0}^{M-1}{\rm e}^{\frac{2\pi {\rm i}}{M}kx}=\frac{1-{\rm e}^{2\pi {\rm i}x}}{1-{\rm e}^{\frac{2\pi {\rm i}}{M}x}}. 
\]
By differentiating this relation we get
\[
\sum_{k=0}^{M-1}k{\rm e}^{\frac{2\pi {\rm i}}{M}kx}
=\frac{-M{\rm e}^{2\pi {\rm i}x}\left( 1-{\rm e}^{\frac{2\pi {\rm i}}{M}x} \right)
+{\rm e}^{\frac{2\pi {\rm i}}{M}x} \left(1-{\rm e}^{2\pi {\rm i}x} \right)}
{\left(1-{\rm e}^{\frac{2\pi {\rm i}}{M}x}\right)^2}
\]
whence
\[
\langle \phi _p|A_f |\phi _q\rangle \!=\!\left\{
\begin{array}{ccc}
\frac{M-1}{2} & {\rm if} & p=q\\[3mm]
\frac{1}{{\rm e}^{\frac{2\pi {\rm i}}{M}(p-q)}-1} & {\rm if} & p\not= q
\end{array} \right. \qquad {\rm in\ the\ case\ }\ \ f(k)\!=\!k.
\]
\subsection{Finite quantum systems}
The study of quantum systems described by finite-dimensional spaces 
was initiated by Weyl \cite{W} and Schwinger \cite{Sch}
and rely upon the discrete Fourier transform.
Let $n$ be a fixed positive integer. The set $\mathbb{Z}_n\times \mathbb{Z}_n\times \mathbb{Z}_n$ considered
together with the multiplication law
\[
(\theta ,\alpha ,\beta )(a',\alpha ',\beta ')
=(\theta +\theta '+\beta \alpha ', \alpha +\alpha ',\beta +\beta ')
\]
where all sums are modulo $n$, is a group. This group of order $n^3$ is regarded as a discrete version
of the Heisenberg group \cite{ST}.\\[5mm]
In any $n$-dimensional Hilbert space $\mathcal{H}$ we can define by choosing an orthonormal basis 
$\{ |0\rangle ,|1\rangle ,...,|n-1\rangle \}$ the Weyl operators $A,B:\mathcal{H}\longrightarrow \mathcal{H}$
\[
A|j\rangle =|j-1\rangle , \qquad B|j\rangle ={\rm e}^{\frac{2\pi {\rm i}}{n}j}|j\rangle 
\]
satisfying the relation
\[
A^\alpha  B^\beta ={\rm e}^{\frac{2\pi {\rm i}}{n}\alpha \beta }\, B^\beta A^\alpha .
\]
The mapping 
\[
(\theta ,\alpha ,\beta )\mapsto {\rm e}^{\frac{2\pi {\rm i}}{n}\theta }\, A^\alpha B^\beta 
\]
defines a unitary irreducible representation of the discrete Heisenberg group in $\mathcal{H}$ and
for any vector $|v\rangle =\sum_{k=0}^{n-1}\nu_k|k\rangle $ we have
\[
{\rm e}^{\frac{2\pi {\rm i}}{n}\theta }\, A^\alpha B^\beta |v\rangle =
{\rm e}^{\frac{2\pi {\rm i}}{n}(\theta +\alpha \beta )}
\sum_{k=0}^{n-1}{\rm e}^{\frac{2\pi {\rm i}}{n}\beta k} \nu_{k+\alpha }|k\rangle 
\]

If we multiply the vectors $ |u_1\rangle ,|u_2\rangle ,...,|u_m\rangle $ of a frame by arbitrary phase
factors we get a new frame 
$ {\rm e}^{{\rm i}\theta _1}|u_1\rangle ,{\rm e}^{{\rm i}\theta _2}|u_2\rangle 
,...,{\rm e}^{{\rm i}\theta _m}|u_m\rangle $.

By choosing a unit vector $|u\rangle =\sum_{k=0}^{n-1}\mu_k|k\rangle $ with stationary group
$G_u\!=\!\mathbb{Z}_n\times \{ 0\}\times \{ 0\}$ and neglecting the phase factors we get the frame \cite{Z}
\begin{equation}
\left\{ \left. |\alpha ,\beta \rangle =
\sum_{k=0}^{n-1}{\rm e}^{\frac{2\pi {\rm i}}{n}\beta k} \mu_{k+\alpha }|k\rangle \ \right|\ \ 
(\alpha ,\beta )\in \mathbb{Z}_n\times \mathbb{Z}_n\ \right\}
\end{equation}
and the resolution of identity
\begin{equation}
\frac{1}{n}\sum_{\alpha ,\beta =0}^{n-1}|\alpha ,\beta \rangle \langle \alpha ,\beta |=\mathbb{I}_\mathcal{H}.
\end{equation}
In the case $n=3$, by choosing 
$|u\rangle =\frac{1}{\sqrt{2}}\, |0\rangle +\frac{1}{\sqrt{2}}\, |1\rangle $ we obtain the frame
\[ 
\fl 
\begin{array}{lll}
|0,0\rangle \!=\! \frac{1}{\sqrt{2}} |0\rangle + \frac{1}{\sqrt{2}}\, |1\rangle \quad &
|0,1\rangle \!=\!\frac{1}{\sqrt{2}}\, |0\rangle +\frac{1}{\sqrt{2}}\varepsilon \, |1\rangle \quad &
|0,2\rangle \!=\!\frac{1}{\sqrt{2}}\, |0\rangle +\frac{1}{\sqrt{2}}\varepsilon ^2\, |1\rangle \\[3mm]
|1,0\rangle \!=\!\frac{1}{\sqrt{2}}\, |0\rangle +\frac{1}{\sqrt{2}}\, |2\rangle \quad &
|1,1\rangle \!=\!\frac{1}{\sqrt{2}}\, |0\rangle +\frac{1}{\sqrt{2}}\varepsilon ^2\, |2\rangle \quad &
|1,2\rangle \!=\!\frac{1}{\sqrt{2}}\, |0\rangle +\frac{1}{\sqrt{2}}\varepsilon \, |2\rangle \\[3mm]
|2,0\rangle \!=\!\frac{1}{\sqrt{2}}\, |1\rangle +\frac{1}{\sqrt{2}}\, |2\rangle \quad &
|2,1\rangle \!=\!\frac{1}{\sqrt{2}}\varepsilon \, |1\rangle \!+\!\frac{1}{\sqrt{2}}\varepsilon ^2\, |2\rangle \quad &
|2,2\rangle \!=\!\frac{1}{\sqrt{2}}\varepsilon ^2\, |1\rangle \!+\!\frac{1}{\sqrt{2}}\varepsilon \, |2\rangle 
\end{array}
\]
where $\varepsilon ={\rm e}^{\frac{2\pi {\rm i}}{3}}$.\\[5mm]
The set $\mathbb{Z}_n\times \mathbb{Z}_n$ can be regarded as a finite version of the phase space,
and to each {\it classical observable} $f:\mathbb{Z}_n\times \mathbb{Z}_n\longrightarrow \mathbb{R}$ 
we associate the linear operator
\begin{equation}
A_f:\mathcal{H}\longrightarrow \mathcal{H},\qquad 
A_f=\frac{1}{n}\sum_{\alpha ,\beta =0}^{n-1}f(\alpha ,\beta )\,|\alpha ,\beta \rangle \langle \alpha ,\beta |.
\end{equation}
For example, in the case $n=2$ by starting from 
$|u\rangle =\frac{3}{5}\, |0\rangle +\frac{4}{5}\, |1\rangle $ we get the frame
\[
\begin{array}{ll}
|0,0\rangle =\frac{3}{5}\, |0\rangle +\frac{4}{5}\, |1\rangle ,\qquad &
|0,1\rangle =\frac{3}{5}\, |0\rangle -\frac{4}{5}\, |1\rangle \\[3mm]
|1,0\rangle =\frac{4}{5}\, |0\rangle +\frac{3}{5}\, |1\rangle ,\qquad &
|1,1\rangle =\frac{4}{5}\, |0\rangle -\frac{3}{5}\, |1\rangle 
\end{array}
\]
and to each function $f:\mathbb{Z}_2\times \mathbb{Z}_2\longrightarrow \mathbb{R}$ we associate the operator
\[ 
\begin{array}{rl}
A_f & =\frac{1}{2}\sum_{\alpha ,\beta =0}^{1}f(\alpha ,\beta )\,|\alpha ,\beta \rangle \langle \alpha ,\beta |\\[5mm]
 & =\frac{1}{50}\left\{ f(0,0)
\left(
\begin{array}{cc}
9 & 12\\
12 & 16
\end{array}
\right) 
+f(1,0)
\left(
\begin{array}{cc}
16 & 12\\
12 & 9
\end{array}
\right)\right. \\[5mm]
 & \qquad \left. +f(0,1)
\left(
\begin{array}{cc}
9 & -12\\
-12 & 16
\end{array}
\right)
+f(1,1)
\left(
\begin{array}{cc}
16 & -12\\
-12 & 9
\end{array}
\right)
\right\}.
\end{array}
\]
We have $\langle 0,0|0,0\rangle =\langle 1,0|1,0\rangle =\langle 0,1|0,1\rangle =\langle 1,1|1,1\rangle =1$ and
\[
\begin{array}{ll}
\langle 0,0|0,1\rangle =-\frac{7}{25}\qquad & \langle 0,1|1,0\rangle =0\\[3mm]
\langle 0,0|1,0\rangle =\frac{24}{25} & \langle 0,1|1,1\rangle =\frac{24}{25}\\[3mm]
\langle 0,0|1,1\rangle =0 & \langle 1,0|1,1\rangle =\frac{7}{25}
\end{array}
\]
and the lower symbol is
\[  
\begin{array}{l}
\langle 0,0|A_f|0,0\rangle 
=\frac{1}{2}\left\{ f(0,0)+f(0,1)\left(\frac{7}{25}\right)^2+f(1,0)\left(\frac{24}{25}\right)^2\right\}\\[5mm]
\langle 0,1|A_f|0,1\rangle 
=\frac{1}{2}\left\{ f(0,0)\left(\frac{7}{25}\right)^2+f(0,1)+f(1,1)\left(\frac{24}{25}\right)^2\right\}\\[5mm]
\langle 1,0|A_f|1,0\rangle 
=\frac{1}{2}\left\{ f(0,0)\left(\frac{24}{25}\right)^2+f(1,0)+f(1,1)\left(\frac{7}{25}\right)^2\right\}\\[5mm]
\langle 1,1|A_f|1,1\rangle 
=\frac{1}{2}\left\{ f(0,1)\left(\frac{24}{25}\right)^2+f(1,0)\left(\frac{7}{25}\right)^2+f(1,1)\right\}.
\end{array}
\]
One can remark that the lower symbols corresponding to the classical observables
we have to analyze depend on the fiducial vector. Therefore, the fiducial vector we use
must be a privileged one, for example, a kind of fundamental state.
We should also notice the way the values of the observables are ``redistributed'' along the probability distribution. 
\subsection{An application of the frame quantization to crystals}
The set $\mathbb{Z}\times \mathbb{Z}$ can be regarded as a mathematical model 
for a two-dimensional crystal. By imposing the cyclic boundary condition, the 
space $\mathcal{E}\!=\!l^2(\mathbb{Z}_N\times \mathbb{Z}_N)$ and the operator
\begin{equation}
\begin{array}{rl}
H:\mathcal{E}\longrightarrow \mathcal{E},\quad 
(H\psi )(n_1,n_2)\! & =\!\psi (n_1\!+\!1,n_2)\!+\!\psi (n_1\!-\!1,n_2)\\[3mm] 
 & \ \ +\!\psi (n_1,n_2\!+\!1)\!+\!\psi (n_1,n_2\!-\!1)
\end{array}
\end{equation}
allow one to describe the electron evolution inside the crystal in the tight binding approximation \cite{Mo}.
For any $k=(k_1,k_2)\in \mathbb{Z}_N\times \mathbb{Z}_N$,  the function 
\begin{equation}
\psi _k:\mathbb{Z}_N\times \mathbb{Z}_N\longrightarrow \mathbb{C},\qquad 
\psi _k(n_1,n_2)={\rm e}^{\frac{2\pi {\rm i}}{N}(k_1n_1+k_2n_2)}
\end{equation}
is an eigenfunction of $H$ corresponding to the eigenvalue
\begin{equation}
E_k={\rm e}^{\frac{2\pi {\rm i}}{N}k_1}+{\rm e}^{-\frac{2\pi {\rm i}}{N}k_1}
+{\rm e}^{\frac{2\pi {\rm i}}{N}k_2}+{\rm e}^{-\frac{2\pi {\rm i}}{N}k_2}
=2\, \cos \frac{2\pi }{N}k_1+2\, \cos \frac{2\pi }{N}k_2,
\end{equation}
that is, 
\[
H\psi _k=E_k\psi _k.
\]
One can remark that 
\[
E_k=\sum_{(n_1,n_2)\in \mathcal{C}}\psi _k(n_1,n_2)
\]
where $\mathcal{C}$ is the cluster
\[
\mathcal{C}=\{ (1,0),(-1,0),(0,1),(0,-1)\} \subset \mathbb{Z}_N\times \mathbb{Z}_N.
\]
The Hilbert space $l^2(\mathcal{C})$ can be identified with the subspace 
\[
\mathcal{H}=\{ \ \varphi :\mathbb{Z}_N\times \mathbb{Z}_N\longrightarrow \mathbb{C} \ |\ \ 
\varphi (n_1,n_2)=0\ \ {\rm for}\ \ (n_1,n_2)\not\in \mathcal{C}\ \}.
\]
The $N^2$ functions  
$\{ \, |\delta _{(n_1,n_2)}\rangle 
=\delta _{(n_1,n_2)}:\mathbb{Z}_N\!\times \!\mathbb{Z}_N\longrightarrow \mathbb{C}\, \}_{n_1,n_2\in \mathbb{Z}_N}$
\[
\delta _{(n_1,n_2)}(n'_1,n'_2)=\left\{
\begin{array}{lll}
1 & {\rm if} & (n'_1,n'_2)=(n_1,n_2)\\
0 & {\rm if} & (n'_1,n'_2)\not=(n_1,n_2)
\end{array} \right.
\]
and the $N^2$ functions 
$\{ \, |\psi _{(k_1,k_2)}\rangle 
=\psi _{(k_1,k_2)}:\mathbb{Z}_N\!\times \!\mathbb{Z}_N\longrightarrow \mathbb{C}\, \}_{k_1,k_2\in \mathbb{Z}_N}$
\begin{equation}
\psi _{(k_1,k_2)}(n_1,n_2)=\frac{1}{N}{\rm e}^{\frac{2\pi {\rm i}}{N}(k_1n_1+k_2n_2)}
\end{equation}
form two orthonormal bases of $\mathcal{E}$ related through the discrete Fourier transform.\\
The orthogonal projector corresponding to $\mathcal{H}$ is 
\[
\pi =\sum_{(n_1,n_2)\in \mathcal{C}}|\delta _{(n_1,n_2)}\rangle \langle \delta _{(n_1,n_2)}|
\]
and in view of theorem 1,  the $N^2$ functions 
$\{ \, |k_1,k_2\rangle :\mathbb{Z}_N\!\times \!\mathbb{Z}_N\longrightarrow \mathbb{C}\, \}_{k_1,k_2\in \mathbb{Z}_N}$
\[ \fl
|k_1,k_2\rangle \!=\!\frac{1}{2}
\sum_{(n_1,n_2)\in \mathcal{C}}|\delta _{(n_1,n_2)}\rangle \langle \delta _{(n_1,n_2)}|\psi_{(k_1,k_2)}\rangle 
\!=\!\frac{1}{2}\sum_{(n_1,n_2)\in \mathcal{C}}
{\rm e}^{\frac{2\pi {\rm i}}{N}(k_1n_1+k_2n_2)}|\delta _{(n_1,n_2)}\rangle 
\]
form a frame in $\mathcal{H}$
\[
\frac{4}{N^2}\sum_{k_1,k_2=0}^{N-1}|k_1,k_2\rangle \langle k_1,k_2|=\mathbb{I}_\mathcal{H}.
\]
They satisfy the relation
\[
\langle k_1,k_2|k'_1,k'_2\rangle =\frac{1}{4}\sum_{(n_1,n_2)\in \mathcal{C}}
{\rm e}^{\frac{2\pi {\rm i}}{N}[(k'_1-k_1)n_1+(k'_2-k_2)n_2]}
\]
\[
\qquad \qquad \qquad =\frac{1}{2}\left[ \cos\frac{2\pi }{N}(k'_1-k_1)+\cos \frac {2\pi }{N}(k'_2-k_2)\right].
\]
To a {\it classical} observable defined by 
$f:\mathbb{Z}_N\times \mathbb{Z}_N\longrightarrow \mathbb{R}$
we associate the linear operator
\begin{equation}
A_f:\mathcal{H}\longrightarrow \mathcal{H},\qquad 
A_f=\frac{4}{N^2}\sum_{k_1,k_2=0}^{N-1}f(k_1,k_2)\, |k_1,k_2\rangle \langle k_1,k_2|
\end{equation}
with the lower symbol
\[
\fl 
\langle k_1,k_2|A_f|k_1,k_2\rangle \!=\!
\frac{1}{N^2}\sum_{k'_1,k'_2=0}^{N-1}f(k'_1,k'_2)\, 
\left[ \cos\frac{2\pi }{N}(k'_1\!-\!k_1)\!+\!\cos \frac {2\pi }{N}(k'_2\!-\!k_2)\right]^2.
\]
In the case of the frame quantization we analyze a classical observable by using a suitable
smaller dimensional subspace. We can increase the resolution of our analysis by choosing a larger
cluster including second order or second and third order neighbours of $(0,0)$.
\subsection{Quantization with finite tight frames overcomplete by one vector}
For each positive integer $n$ we consider in the Euclidean space $\mathbb{R}^{n+1}$ the hyperspace
\[
\mathcal{H}_n=\{ x=(x_0,x_1,...,x_n)\ |\ x_0+x_1+\cdots +x_n=0\ \}.
\]
The orthogonal projector corresponding to $\mathcal{H}_n$ is $\pi :\mathbb{R}^{n+1}\longrightarrow \mathbb{R}^{n+1}$,
\[
\begin{array}{l}
\pi (x_0,x_1,\dots ,x_n)
=\left( \frac{nx_0-x_1-\cdots -x_n}{n+1},\frac{-x_0+nx_1-x_2-\cdots -x_n}{n+1},\dots
\frac{-x_0-\cdots -x_{n-1}+nx_n}{n+1}\right)
\end{array}
\]
and the orthogonal projections of the vectors of the canonical basis
\[
\begin{array}{l}
w_0=\pi (1,0,0,\dots ,0)=\left( \frac{n}{n+1},-\frac{1}{n+1},-\frac{1}{n+1},\dots ,-\frac{1}{n+1} \right)\\[3mm]
w_1=\pi (0,1,0,\dots ,0)=\left( -\frac{1}{n+1},\frac{n}{n+1},-\frac{1}{n+1},\dots ,-\frac{1}{n+1} \right)\\[3mm]
...................................................................................\\[3mm]
w_n=\pi (0,0,\dots ,0,1)=\left( -\frac{1}{n+1},-\frac{1}{n+1},\dots ,-\frac{1}{n+1},\frac{n}{n+1} \right)
\end{array}
\]
have the same norm
\[
||w_0||=||w_1||=\cdots =||w_n||=\sqrt{\frac{n}{n+1}}.
\]
The corresponding normalized vectors 
\[
\begin{array}{l}
|u_0\rangle =\frac{w_0}{||w_0||}
=\left( \sqrt{\frac{n}{n+1}},-\frac{1}{\sqrt{n(n+1)}},-\frac{1}{\sqrt{n(n+1)}},\dots ,-\frac{1}{\sqrt{n(n+1)}}\right)\\[3mm]
|u_1\rangle =\frac{w_1}{||w_1||}
=\left(-\frac{1}{\sqrt{n(n+1)}}, \sqrt{\frac{n}{n+1}},-\frac{1}{\sqrt{n(n+1)}},\dots ,-\frac{1}{\sqrt{n(n+1)}}\right)\\[3mm]
...................................................................................\\[3mm]
|u_n\rangle =\frac{w_n}{||w_n||}
=\left(-\frac{1}{\sqrt{n(n+1)}}, -\frac{1}{\sqrt{n(n+1)}},\dots ,-\frac{1}{\sqrt{n(n+1)}},\sqrt{\frac{n}{n+1}}\right)
\end{array}
\]
form a normalized tight frame
\[
\frac{n}{n+1}\sum_{k=0}^n|u_k\rangle \langle u_k|=\mathbb{I}_{\mathcal{H}_n}
\]
such that
\[
\langle u_k|u_j\rangle =-\frac{1}{n}\qquad {\rm for}\ k\not=j.
\]
To each function $f:\{ 0,1,\dots ,n\}\longrightarrow \mathbb{R}$ we associate the linear operator
\[
A_f:\mathcal{H}_n\longrightarrow \mathcal{H}_n,\qquad 
A_f=\frac{n}{n+1}\sum_{k=0}^nf(k)\, |u_k\rangle \langle u_k|.
\]
The corresponding lower symbol is the function $\check f_n:\{ 0,1,\dots ,n\}\longrightarrow \mathbb{R}$,
\[
\check f_n(j)=\langle u_j|A_f|u_j\rangle =\frac{n-1}{n}f(j)+\frac{1}{n(n+1)}\sum_{k=0}^nf(k)
\]
and if $f:\{ 0,1,2,\dots \}\longrightarrow \mathbb{R}$ is a bounded function then we have
\[
\lim_{n\rightarrow \infty }\check f_n(j)=f(j), \qquad {\rm for\ any}\ j\!\in \!\{ 0,1,2, \dots \},
\]
as expected from the general results given in subsection \ref{classlim}.
More than that, if $f,g:\{ 0,1,2,\dots \}\longrightarrow \mathbb{R}$ are two bounded functions then
\[ \fl
\begin{array}{l}
\langle u_j|[A_f,A_g]|u_j\rangle =
\frac{n^2}{(n+1)^2}\sum_{k=0}^n\sum_{l=0}^n(f(k)g(l)-f(l)g(k))\, 
\langle u_j|u_k\rangle \, \langle u_k|u_l\rangle \, \langle u_l|u_j\rangle \\[3mm]
\qquad \qquad \qquad \ =-\frac{1}{n(n+1)^2}\sum_{k\not=j}\sum_{l\not=j}(f(k)g(l)-f(l)g(k))
\end{array}
\]
and the lower symbol of the commutator $[A_f,A_g]$ has the property
\[
\lim_{n\rightarrow \infty }\langle u_j|[A_f,A_g]|u_j\rangle =0, \qquad {\rm for\ any}\ j\!\in \!\{ 0,1,2, \dots \}.
\]
%
%
%
%
\section{Conclusions}
In this paper we have presented some elements concerning certain applications of finite frames to crystal/quasicrystal physics
and to quantum physics. In order to achieve these two main objectives and inspired by the analogy with standard coherent states, 
we have introduced the notion of normalized Parseval frame, directly related to the notion of Parseval frame, and analyzed some stochastic aspects.  In particular we have defined two types of ``distances'' , $\eta = r-1$ and $\zeta = 1 - \min_{1\leq i \leq M}\kappa_i$, between frames and orthonormal basis in the superspace. 
For the applications to crystals and quasicrystals, based on the embedding into a superspace defined by a frame, we have analyzed
the subset of the elements which can be represented as a linear combination of frame vectors by using only integer coefficients.
We have identified in this way two important classes of tight frames, namely the periodic frames and the quasiperiodic frames.
We have also presented some convergent sequences of finite frames and an example of continuous deformation of a periodic tetrahedral
frame into an icosahedral quasiperiodic frame. Some of these theoretical considerations seem to be new, and might be regarded as a contribution
to the finite frame theory.

The description of the elements of a vector space based on the use of an overcomplete system is a general method re-discovered several
times in different areas of mathematics, science and engineering . For example, in crystallography there exists an alternative description for the hexagonal
crystals based on the use of an additional axis. We show that the use of a frame leads to a simpler description of atomic positions in a 
diamond type crystal. This leads to a simpler description of the symmetry transformations and of the mathematical objects with physical meaning.
Some of the most important models used in quasicrystal physics can be generated in a unitary way by using the imbedding into a superpace
defined by certain frames. These observations allow a fructuous  interchange of ideas and methods between frame theory and quasicrystal physics.

Finite frame quantization replaces a real function $f$ defined on a finite set by a self-adjoint operator $A_f$, and the eigenvalues of $A_f$
can be regarded as the ``quantum spectrum'' of $f$. We compare $f$ with the mean values of $A_f$ corresponding to the frame vectors,
in the general case and in several particular cases. We have explained the role of the parameter $\zeta$ as a kind of distance of the quantum non-commutative world to the classical commutative one. The notion of normalized Parseval frame and the corresponding quantization of
discrete variable functions is rich of questions which deserve to be thoroughly investigated in the measure that they might  shed light on a better understanding of quantum mechanics and quantization. %
\section*{Acknowledgment}
NC acknowledges the support provided by CNCSIS under the grant IDEI 992 - 31/2007.

\end{document}